\documentclass{report}   
\usepackage{svcon2ex}
\usepackage{epsfig}

\Draft         
\editedby{}
\SetVersionDate{Jan. 21, 2002}

\begin{document}
\pagenumbering{arabic} 
\chapter{Boolean Dynamics with Random Couplings}
\shortchapname{Dynamics in Boolean Networks}       

\chapterauthors{Maximino Aldana\\ 
  Susan Coppersmith\\ 
  Leo P. Kadanoff}
\shortauthname{M. Aldana and S. Coppersmith and L. Kadanoff}

 

\begin{abstract}
  This paper reviews a class of generic dissipative dynamical systems
  called $N$-$K$ models.  In these models, the dynamics of $N$
  elements, defined as Boolean variables, develop step by step,
  clocked by a discrete time variable.  Each of the $N$ Boolean
  elements at a given time is given a value which depends upon $K$
  elements in the previous time step.
  
  We review the work of many authors on the behavior of the models,
  looking particularly at the structure and lengths of their cycles,
  the sizes of their basins of attraction, and the flow of information
  through the systems. In the limit of infinite $N$, there is a phase
  transition between a chaotic and an ordered phase, with a critical
  phase in between.
  
  We argue that the behavior of this system depends significantly on
  the topology of the network connections.  If the elements are placed
  upon a lattice with dimension $d$, the system shows correlations
  related to the standard percolation or directed percolation phase
  transition on such a lattice.  On the other hand, a very different
  behavior is seen in the {\em Kauffman net} in which all spins are
  equally likely to be coupled to a given spin.  In this situation,
  coupling loops are mostly suppressed, and the behavior of the system
  is much more like that of a mean field theory.
  
  We also describe possible applications of the models to, for
  example, genetic networks, cell differentiation, evolution,
  democracy in social systems and neural networks.
\end{abstract}

\nocite{*}

\section{Introduction}

In this review, we describe the dynamics of a set of $N$ variables, or
elements, which each have two possible values (say 0 and 1). These
elements interact with each other according to some given interaction
rules, specified through a set of Boolean coupling functions that
determine the variables at the next time-step, and thereby give the
dynamics of the system.  Such a discrete stepping of a set of Boolean
variables, also known in general terms as a {\em Boolean network}, is
of potential interest in several different fields, ranging from gene
regulation and control, to modeling democracy and social organization,
to understanding the behavior of glassy materials.

The models were originally studied primarily for their biological
interest, specifically by Stuart Kauffman who introduced the so-called
$N$-$K$ \emph{model} in the context of gene expression and fitness
landscapes in 1969 (\cite{69.1,74.1,95.4,93.1,90.3,84.1}).  Since
Kauffman's original work, the scientific community has found a broad
spectrum of applicability of these models.  Specific biological
problems studied include cell differentiation (\cite{00.4}), immune
response (\cite{89.1}), evolution (\cite{98.5,98.10,00.8,94.1}),
regulatory networks (\cite{00.5}) and neural networks
(\cite{90.1,87.4,88.11,00.5}).  In the first two examples, the basic
binary element might be a chemical compound, while in the last it
might be the state of firing of a neuron.  A computer scientist might
study a similar set of models, calling the basic elements {\em gates},
and be thinking about the logic of computer design (\cite{81.1,95.5})
or optimization (\cite{98.3, 94.3}).  Earlier work in the mathematical
literature (\cite{60.1,53.1}) studied
\emph{random mapping models}, which are a subset of the models
introduced by Kauffman.  This same kind of problem has also drawn
considerable attention from physicists interested in the development
of chaos (\cite{98.8,98.9,97.3,97.7,97.10,96.1,96.4,95.5}) and also in
problems associated with glassy and disordered materials
(\cite{86.2,86.3,87.5,87.6}).  In these examples, the Boolean element
might be an atomic spin or the state of excitation of a molecule.

In some sense, the type of Boolean networks introduced by Kauffman can
be considered as a prototype of generic dynamical system, as they
present chaotic as well as regular behavior and many other typical
structures of dynamical systems. In the thermodynamic limit
$N\rightarrow \infty$, there can be ``phase transitions''
characterized by a critical line dividing chaotic from regular regions
of state space. The study of the behavior of the system at and near
the phase transitions, which are attained by changing the
model-parameters, has been a very major concern.

As we shall describe in more detail below, these models are often
studied in a version in which the couplings among the Boolean
variables are picked randomly from some sort of ensemble. In fact,
they are often called $N$-$K$ models because each of the $N$ elements
composing the system, interact with exactly $K$ others (randomly
chosen).  In addition, their coupling functions are usually picked at
random from the space of all possible functions of $K$ Boolean
variables.  Clearly this is a simplification of real systems as there
is no particular problem which has such a generically chosen coupling.
All real physical or biological problems have very specific couplings
determined by the basic structure of the system in hand.  However, in
many cases the coupling structure of the system is very complex and
completely unknown.  In those cases the only option is to study the
generic properties of generic couplings.  One can then hope that the
particular situation has as its most important properties ones which
it shares with generic systems.

Another simplification is the binary nature of the variables under
study.  Nevertheless, many systems have important changes in behavior
when ``threshold'' values of the dynamical variables are reached (e.g.
the synapses firing potential of a neuron, or the activation potential
of a given chemical reaction in a metabolic network). In those cases,
even though the variables may vary continuously, the binary approach
is very suitable, representing the above-below threshold state of the
variables.  The Boolean case is particularly favorable for the study
of generic behavior.  If one were to study a continuum, one would have
to average the couplings over some rather complicated function space.
For the Booleans, the function space is just a list of the different
possible Boolean functions of Boolean variables.  Since the space is
enumerable, there is a very natural measure in the space. The averages
needed for analytic work or simulations are direct and easy to define.

In addition to its application, the study of generic systems is of
mathematical interest in and for itself.

\subsection{Structure of Models}

Any model of a Boolean net starts from $N$ elements $\{\sigma_1$,
$\sigma_2$, $\dots$, $\sigma_N\}$, each of which is a binary variable
$\sigma_i \ \in \ \{0,1\}$, $i = 1, 2, \dots , N$. In the time
stepping, each of these Boolean elements is given by a function of the
other elements. More precisely, the value of $\sigma_i$ at time $t+1$
is determined by the value of its $K_i$ \emph{controlling elements}
$\sigma_{j_1(i)}$, $\sigma_{j_2(i)}$, $\dots$, $\sigma_{j_{K_i}(i)}$
at time $t$. In symbols,
\begin{equation}
\sigma_i(t+1)  = f_i(\sigma_{j_1(i)}(t),\sigma_{j_2(i)}(t),
...,\sigma_{j_{K_i}(i)}(t)),
\label{m1}
\end{equation}
where $f_i$ is a Boolean function associated with the $i^{th}$ element
that depends on $K_i$ arguments.  To establish completely the model it
is necessary to specify:

\begin{itemize}
\item the \emph{connectivity} $K_i$ of each element, namely, how many
  variables will influence the value of every $\sigma_i$;
\item the \emph{linkages} (or \emph{couplings}) of each element, which
  is the particular set of variables $\sigma_{j_1(i)},
  \sigma_{j_2(i)}, \dots , \sigma_{j_{K_i}(i)}$ on which the element
  $\sigma_i$ depends, and
\item the \emph{evolution rule} of each element, which is the Boolean
  function $f_i$ determining the value of $\sigma_i(t+1)$ from the
  values of the linkages $\sigma_{j_1(i)}(t)$, $\sigma_{j_2(i)}(t)$,
  $\dots$, $\sigma_{j_{K_i}(i)}(t)$.
\end{itemize}
 
Once these quantities have been specified, equation (\ref{m1}) fully
determines the dynamics of the system. In the most general case, the
connectivities $K_i$ may vary from one element to another. However,
throughout this work we will consider only the case in which the
connectivity is the same for all the nodes: $K_i = K$, $i = 1,
2,\dots, N$. In doing so, it is possible to talk about \emph{the}
connectivity $K$ of the whole system, which is an integer parameter by
definition. It is worth mentioning though that when $K_i$ varies from
one element to another, the important parameter is the \emph{mean}
connectivity of the system, $\langle K \rangle$, defined as
\[
\langle K \rangle = \frac{1}{N}\sum^N_{i=1}K_i  .
\]
In this way, the mean connectivity might acquire non-integer values.
Scale-free networks (\cite{01.4,01.5}), which have a very broad
(power-law) distribution of $K_i$, can also be defined and
characterized.

\begin{figure}[t]
\centering
\psfig{file=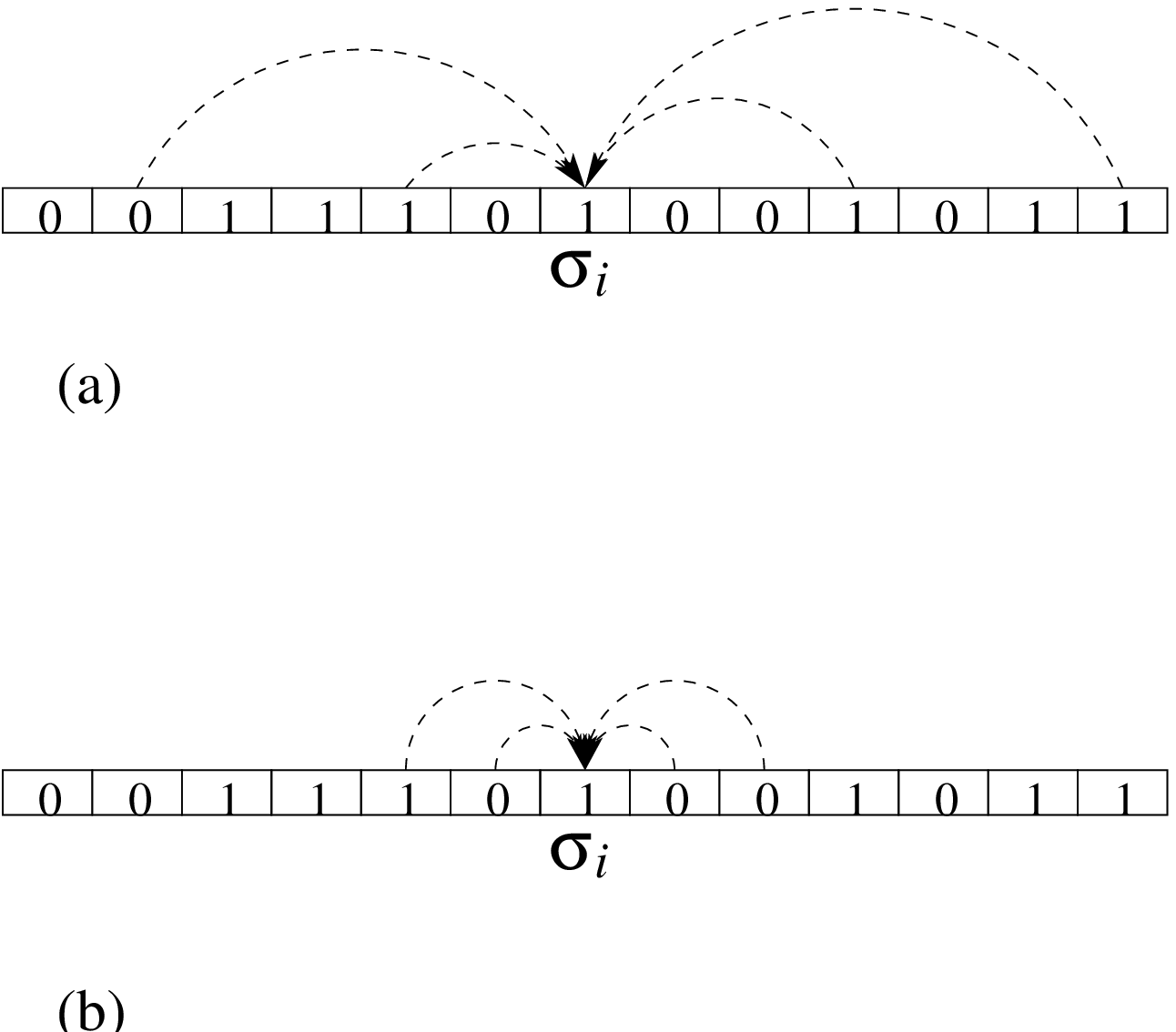,clip=,height=3in}
\caption[]{ \small The different kinds of linkages in a one 
  dimensional system.  (a) In the Kauffman net the linkages of every
  element $\sigma_i$ are chosen at random among all the other elements
  $\sigma_1$ $\dots$ $\sigma_N$. (b) In a completely ordered lattice,
  the linkages are chosen according to the geometry of the space. In
  the case illustrated in this figure, $\sigma_i$ is linked to its
  first and second nearest neighbors.}
\label{first}
\end{figure}

Of fundamental importance is the way the linkages are assigned to the
elements, as the dynamics of the system both qualitatively and
quantitatively depend strongly on this assignment.  Throughout this
paper, we distinguish between two different kinds of assignment: In a
{\em lattice assignment} all the bonds are arranged on some regular
lattice. For example, the $K$ control elements $\sigma_{j_1(i)}$,
$\sigma_{j_2(i)}$, $\dots$, $\sigma_{j_K(i)}$ may be picked from among
the $2 d$ nearest neighbors on a $d$ dimensional hyper-cubic lattice.
Alternatively, in a {\em uniform assignment} each and every element
has an equal chance of appearing in this list.  We shall call a
Boolean system with such a uniform assignment a {\em Kauffman net}.
(See Fig. \ref{first}.)

%
%
%
%
%
%
%
%

Of course, intermediate cases are possible, for example, one may
consider systems with some linkages to far-away elements and others to
neighboring elements.
\emph{Small-world networks} (\cite{01.4}) are of this type.
 
For convenience, we will denote the whole set of Boolean elements
$\{\sigma_1(t)$, $\sigma_2(t),\dots,$ $\sigma_N(t)\}$ by the symbol
$\Sigma_t$:
\begin{equation}
\Sigma_t = \{\sigma_1(t),\sigma_2(t),\dots,\sigma_N(t)\} ;
\end{equation}
$\Sigma_t$ represents then the state of the system at time $t$.  We
can think of $\Sigma_t$ as an integer number which is the base-10
representation of the binary chain
$\{\sigma_1(t),\sigma_2(t),\dots,\sigma_N(t)\}$.  Since every variable
$\sigma_i$ has only two possible values, 0 and 1, the number
of all the possible configurations is $\Omega=2^N$, so that $\Sigma_t$
can be thought of as an integer satisfying $0 \leq \Sigma_t < 2^N$.
This collection of integers is the base-10 representation of the
\emph{state space} of the system. Although it is not essential for the
understanding of the underlying dynamics of the network, this integer
representation proves to be very useful in the implementation of the
computational algorithms used in numerical simulations (at least for
small values of $N$).

A note of caution is relevant at this point. We should distinguish the
purely Boolean model described in this work from Kauffman's $N$-$K$
landscape model, which provides a description of fitness landscapes by
including a fitness function to be optimized. We are not going to
review on fitness landscapes since abundant literature already exists
on this topic (\cite{01.6,95.4,93.1}).

\subsection{Coupling Functions}

The arguments of the coupling functions $f_i(\sigma_{j_1(i)}$,
$\dots$, $\sigma_{j_K(i)})$ can take on $2^K$ different values. One
specifies the functions by giving, for each of these values of the
arguments, a value to the function.  Therefore there are a total of
\begin{equation}
\aleph = 2^{2^K}
\end{equation}
different possible functions. In Table \ref{boolean_f1} we give two
examples of coupling functions for the case $K=3$. There are $2^3=8$
configurations of the arguments $\sigma_{j_1(i)}$, $\sigma_{j_2(i)}$,
$\sigma_{j_3(i)}$, and for each one of these configurations, the
function $f_i$ can acquire the values $1$ or $0$. For $K=3$ there are
$2^{2^3} = 256$ tables similar to the one shown in Table 1, one for
each Boolean coupling function.  Different tables differ in their
assignments of 0's and 1's.
If we assign a probability or weight to each of these functions, one
gets an {\em ensemble} of possible couplings.

\begin{table}
\center
\begin{tabular}{|ccc|c|c|}
\hline
 & & & Random & Canalizing\\
$\sigma_{j_1}$ & $\sigma_{j_2}$ & $\sigma_{j_3}$ & 
$f(\sigma_{j_1},\sigma_{j_2},\sigma_{j_3})$ &
$f(\sigma_{j_1},\sigma_{j_2},\sigma_{j_3})$\\
\hline
0 & 0 & 0 & 0 & 1\\
0 & 0 & 1 & 1 & 1\\
0 & 1 & 0 & 1 & 1\\
0 & 1 & 1 & 0 & 1\\
1 & 0 & 0 & 1 & 0\\
1 & 0 & 1 & 0 & 1\\
1 & 1 & 0 & 1 & 0\\
1 & 1 & 1 & 1 & 0\\
\hline
\end{tabular}
\caption{Illustration of two Boolean functions of three arguments.
The first function is a particular random function, whereas the second
one is a canalizing function of the first argument $\sigma_1$. When
this argument is 0, the output of the function is always 1, while if
$\sigma_1=1$, the output can be either 0 or 1.}
\label{boolean_f1}
\end{table}

Possible ensenble choices abound.  One ensemble used extensively by
Kauffman and others is the {\em uniform distribution} in which all
functions are weighted equally.  Alternatively, a {\em magnetization
bias}\footnote{The word magnetization comes from the possibility of
identifying each element with an atomic spin, which is a very small
magnet.} may be applied by weighting the choice of functions with an
outcome $0$ with a probability $p$, and the outcome $1$ with a
probability $1-p$ (see, for example \cite{98.6}). One may also give
different weights to particular types of functions. For example, one
can consider only {\em forcing functions} or {\em canalizing
functions} (\cite{87.2,69.1,84.1}), in which the function's value is
determined when just one of its arguments is given a specific
value. The second function shown in Table \ref{boolean_f1} is a
canalizing function.  Another possibility is to specify the value of
the function in order to simulate the additive properties of neurons
(\cite{00.5,99.6,94.5,90.1,88.11,87.4}).

Here we enumerate some of the coupling functions
occurring for different values of the connectivity $K$.

\begin{itemize}
\item For $K=0$ there are but two functions, corresponding to the two
  possible values of a Boolean variables: tautology $f=1$ and
  contradiction $f=0$.  Together these two functions form a class
  which we might call $\mathcal{A}$.
  
\item For $K=1$, in addition to the class $\mathcal{A}$, there exists
  another class $\mathcal B$ in which $f(\sigma)$ can take on the
  value $\sigma$, called identity, and the value $\neg \sigma$, called
  negation.  Thus there are a total of four functions, represented as
  columns in Table \ref{boolean_f2}.

\begin{table}
\center
\begin{tabular}{|c|c|c||c|c|}
\hline
 & \multicolumn{2}{c||}{Class $\mathcal{A}$} & 
\multicolumn{2}{c|}{Class $\mathcal{B}$}\\
\cline{2-5}
$\sigma$ & ${\mathcal A}_0$ & ${\mathcal A}_1$ & ${\mathcal B}_I$ & 
${\mathcal B}_N$\\ 
\hline
0 & 0 & 1 & 0 & 1\\
1 & 0 & 1 & 1 & 0\\
\hline
\end{tabular}
\caption{Boolean functions for $K=1$. The first two functions 
form the class $\mathcal A$ of constant functions, ${\mathcal
A}_0(\sigma)=0$ and ${\mathcal A}_1(\sigma)=1$.  The other two
functions form class $\mathcal B$ which consist in identity ${\mathcal
B}_I(\sigma)=\sigma$ and negation ${\mathcal
B}_N(\sigma)=\neg\sigma$}.
\label{boolean_f2}
\end{table}

\item The situation for $K=2$ has been particularly carefully studied.
  Here there are four classes of functions $f(\sigma_1,\sigma_2)$
  (\cite{93.3,01.2}).  Each class is invariant under making the
  interchange $0 \leftrightarrow 1$ in either arguments or value of
  $f$.  The classes are ${\mathcal A}$ (two constant functions),
  ${\mathcal B}_1$ (four canalizing functions which depend on one
  argument), ${\mathcal B}_2$ (eight canalizing functions which depend
  on two arguments), ${\mathcal C}$, (two non-canalizing functions).
  These functions are explicitly shown in Table \ref{boolean_f3}.

\begin{table}
\center
\begin{tabular}{|cc|c|c||c|c|c|c||c|c|c|c|c|c|c|c||c|c|}
\hline
$\sigma_{j_1}$ & $\sigma_{j_2}$ & 
\multicolumn{2}{c||}{Class ${\mathcal A}$} &
\multicolumn{4}{c||}{Class ${\mathcal B}_1$} &
\multicolumn{8}{c||}{Class ${\mathcal B}_2$} & 
\multicolumn{2}{c|}{Class ${\mathcal C}$} \\
\hline
0&0&1&0&0&1&0&1&1&0&0&0&0&1&1&1&1&0\\
0&1&1&0&0&1&1&0&0&1&0&0&1&0&1&1&0&1\\
1&0&1&0&1&0&0&1&0&0&1&0&1&1&0&1&0&1\\
1&1&1&0&1&0&1&0&0&0&0&1&1&1&1&0&1&0\\
\hline
\end{tabular}
\caption{Boolean functions for the case $K=2$. The 
16 functions can be arranged in four different classes which differ
in their symmetry properties (see text).}
\label{boolean_f3}
\end{table}

\end{itemize}

Several calculations have been done by giving different weights to the
different classes (see for example \cite{95.5,87.2}) .

\subsection{The Updates}

Once the linkages and the $f_i$'s are given, one is said to have
defined a {\em realization}\/ of the model.  Given the realization,
one can define a dynamics by using equation (\ref{m1}) to update all
the elements at the same time.  This is called a \emph{synchronous}
update.  In this paper, we assume a synchronous update unless stated
otherwise.  Alternatively, one may have a \emph{serial} model in which
one updates only one element at a time.  This element may be picked at
random or by some predefined ordering scheme.

Additional choices must be made.  One can:

\begin{enumerate}
\item Keep the same realization through all time.  We then have a
  \emph{quenched} model.
  
\item Pick an entirely new realization after each time step.  The
  model is then said to be {\em annealed}.\footnote{These terms have
    been borrowed from the physics of alloys in which something which
    is cooled quickly so that it cannot change its configuration is
    said be be quenched, and something which is held at a high
    temperature for a long time so that it can respond to its
    environment is described as annealed. Hence these terms are
    applied to situations in which one wants to distinguish between
    problems with fixed versus changing interactions.  }
\item Employ a {\em genetic algorithm} in
  which the system slowly modifies its realization so as to approach a
  predefined goal (\cite{00.8,99.9,94.3}).

\item Intermediate choices are also possible (\cite{94.2,96.3}).
\end{enumerate}

Almost always, we shall regard the {\em real system} as one which
updates synchronously and which is quenched so that the interactions
are fixed for all time. The annealed and sequential models will be
regarded as approximations which can provide clues to the behavior of
this ``real system''.  The quenched model has time-independent
dynamics describing motion within the state space of size $\Omega
=2^N$.  One iterates the model through time by using equation
(\ref{m1}) and thereby obtains a dynamics for the system. Each of the
$\Omega$ different initial conditions will generate a motion, which
will eventually fall into a cyclical behavior.

\subsection{Symmetry Properties}

Typically, each individual realization of these models shows little or
no symmetry.  However, the average over realizations has quite a large
symmetry group, and the symmetry is important to model behavior.  For
example, the random mapping model (\cite{60.1}), which is the
$K\rightarrow\infty$ limit of the $N$-$K$ model of the Kauffman net,
has a full symmetry under the interchange of all states forming the
state space.  For finite values of $K$, the Kauffman net is symmetric
under the interchange of any two basic elements.  One can also have a
symmetry under the interchange of the two values of each individual
element if one chooses the couplings at random, or with the
appropriate symmetry.  One can use dynamics that have reversal
symmetry (\cite{60.1,01.2,01.3}), and that choice will have a profound
effect upon the structure of the cycles.

\subsection{Outline of Paper}

To define fully the object of study, one must describe the dynamical
process and the class of realizations that one wishes to study.  For
example, one can fix $N$ and $K$, and study all of the properties of
all realizations of frozen systems with those values.  One might pick
a realization at random among all possible linkages and functions,
develop its properties.  Then one would pick another realization, and
study that.  Many such steps would give us the properties of the
frozen system averaged over realizations with a given $N$ and $K$.
What properties might we wish to study?

In the next chapter, we describe the gross information transfer
through the system by describing how the system will respond to a
change in initial data or couplings.  There are three different {\em
phases} that have qualitatively different information transfer
properties.  We argue that the Kauffman net, which can transfer
information from any element to any other, is qualitatively different
from lattice systems, in which the information transfer occurs through
a $d$-dimensional space.  We argue that the information transfer on
the lattice is qualitatively, and even quantitatively, similar to the
kind of information flow studied in percolation problems.

Chapter 3 is concerned with the temporal recurrence in the network as
reflected in the statistical properties of its cycles.  Here we base
our arguments upon a discussion of two important limiting cases,
$K=1$, and very large $K$.  The first case is dominated by situations
in which there are a few short linkage loops.  In the second, the
Kauffman net shows a behavior which can be analyzed by comparison with
a random walk through the state space.  The distribution of cycle
lengths is qualitatively different from any of the quantities that are
commonly studied in percolation problems. So the results
characterizing the cycles are substantially different from the
behaviors usually studied in phase transition problems. We argue in
addition that the cycles of Kauffman nets and of networks on
$d$-dimensional lattices differ substantially.

\section{Information Flow}
\subsection{Response to Changes}

The first thing to study in an $N$-$K$ model is its response to
changes.  This response is important because the actual values of the
elements often do not matter at all.  If, for example, we pick the
functions $f_i$ at random among the class of all Boolean functions of
$K$ variables, then the ensemble is invariant under flipping the value
of the $i$th element. In that case, only changes matter, not values.

In computer studies, such changes can be followed quite simply.  One
follows a few different time developments of systems that are
identical except for a small number of selected changes in the
coupling functions or initial data, and sees how the differences
between the configurations change in time.  One can do this for two
configurations or for many, studying pair-wise differences between
states, or things which remain identical across all the
time-developments studied.

\subsubsection{Hamming Distance and Divergence of Orbits}

For simplicity, imagine starting out from two different possible
initial states:
\begin{equation}
\Sigma_0 = \{\sigma_1(0),\sigma_2(0),\ldots,\sigma_N(0)\}  \quad
\tilde{\Sigma}_0 =
\{\tilde{\sigma}_1(0),\tilde{\sigma}_2(0),\ldots,\tilde{\sigma}_N(0)\}
\end{equation}
which differ in the values of a few elements.  One can then observe
the time-development of these configurations under the same dynamics,
finding, for example, the distance $D(t)$ between the configurations as a
function of time 
\begin{equation} D(t) = \sum_{i \ = 1}^N \bigl(\sigma_i(t)
-\tilde{\sigma}_i(t)\bigr)^2 .
\label{dis}
\end{equation} 
If the transfer of information in the system is localized, this
distance will never grow very large.  If however, the system is
sufficiently chaotic so that information may be transferred over the
entire system, then in the limit of large $N$ this Hamming distance can
diverge for large times.

Another interesting measure is the normalized overlap between
configurations, $a(t)$, defined as 
\begin{equation} a(t) = 1 - N^{- 1}D(t) .
\label{overdef}
\end{equation}
One will wish to know whether $a$ goes to unity or a lesser value as
$t\rightarrow\infty$.  If the overlap always goes to unity,
independently of the starting states, then the system cannot retain a
nonzero fraction of the information contained in its starting
configuration. Alternatively, when $a(\infty)$ is less than unity, the
system ``remembers'' a nonzero fraction of its input data. 
 
\subsubsection{Response to Damage}

So far, we have considered the system's response to changes of the
initial data.  One can also attack the quenched problem by considering
two systems, each with the same initial data, but with a few of the
linkages varied.  Then one can ask: given such ``damage'' to the
system, how much do the subsequent states of the system vary?  Do they
become more and more alike or do they diverge?  What is the likelihood
of such a divergence?

These considerations of {\em robustness}---both to damage and to
variation in initial data---are very important for the evaluation of
the effectiveness of a network, either for computations or as part of
a biological system.  There have been fewer studies of the effect of
damage than that of initial data.  Usually the two types of robustness
occur together (\cite{00.2,97.6}). Damage has been studied for
itself (\cite{94.3, 88.9}).

\subsection{Percolation and Phase Behavior}

\subsubsection{Percolation of Information} 
\label{percolateI}

In the limiting case in which $N$ approaches infinity, the different
types of $N$-$K$ models all show three different kinds of phases,
depending upon the form of information transfer in the system. If the
time development transfers information to a number of elements that
grows exponentially in time, the system is said to be in a {\em
chaotic phase}.  Typically, this behavior occurs for larger values of
$K$, up to and including $K=N$.
If, on the other hand, a change in the initial data typically
propagates to only a finite number of other elements, the system is
said to be in a {\em frozen phase}.  This behavior will arise for
smaller values of $K$, most especially $K=0$, and usually $K=1$.
There is an intermediate situation in which information typically
flows to more and more elements as time goes on, but this number
increases only algebraically. This situation is described as a {\em
critical phase}.

When the linkages and the hopping among configurations are
sufficiently random, one can easily perform a quite precise
calculation of the boundary which separates these phases.  Imagine
starting with a state $\Sigma_0$, containing a very large number, $N$,
of Boolean elements, picked at random.  Imagine further another
configuration $\tilde{\Sigma}_0$ in which the vast majority of the
elements have the same value as in $\Sigma_0$, but nevertheless there
are a large number of elements, picked at random, which are different.
The Hamming distance at time zero, $D(0)$, is the number of changed
elements. Now take the simplest $N$-$K$ system in which all the
linkages and the couplings are picked at random. On average, a change
in a single element will change the argument of $K$ functions, so
there will be $K D(0)$ functions affected. Each of these will have a
probability one half of changing their value.  (The functions after
all are quite random.)  Thus the Hamming distance after the first time
step will be
\[
D(1)=0.5 K D(0) .
\]
If the couplings and connections are sufficiently random, then at the
start of the next step, the newly changed elements and their couplings
will remain quite random.  Then the same sort of equation will apply
in the next time step, and the next. Just so long as the fraction of
changed elements remains small, and the randomness continues, the
Hamming distance will continue to change by a factor of $K/2$ so that
\[
D(t+1)=0.5 KD(t) ,
\]
which then has the solution
\begin{equation}
D(t)= D(0) \exp[t \ln(0.5 K)] .
\label{expgrow} 
\end{equation}
For $K>2$ the number of changed elements will grow exponentially, for
$K~<~2$ it will decay exponentially, and for $K=2$ there will be neither
exponential growth nor decay, and the behavior will be substantially
influenced by fluctuations. Thus, by varying the value of the
connectivity, the system sets down into one of the three following
phases:

\begin{itemize}
\item \emph{Chaotic} ($K>2$), the Hamming distance grows exponentially
  with time.
\item \emph{Frozen} ($K<2$), the Hamming distance decays exponentially
  with time.
\item \emph{Critical} ($K_c=2$), the temporal evolution of the Hamming
  distance is determined mainly by fluctuations.
\end{itemize}

In deriving equation (\ref{expgrow}) we have assumed that the coupling
functions $f_i$ of the system acquire the values $0$ and $1$ with the
same probability $p = 1/2$. Nonetheless, as we will see below, the
chaotic, frozen and critical phases are also present in the more
general case in which the coupling functions $f_i$ evaluate to $0$ and
$1$ with probabilities $p$ and $1-p$ respectively. For a given value
of $p$, there is a critical value $K_c(p)$ of the connectivity
below which the system is in the frozen phase and above which the
chaotic phase is attained.  Conversely, for a given connectivity $K\ge 2$,
a critical value $p_c(K)$ of the probability bias
separates the chaotic and the frozen phases.

%
%
%
%
\begin{figure}[h]
\psfig{file=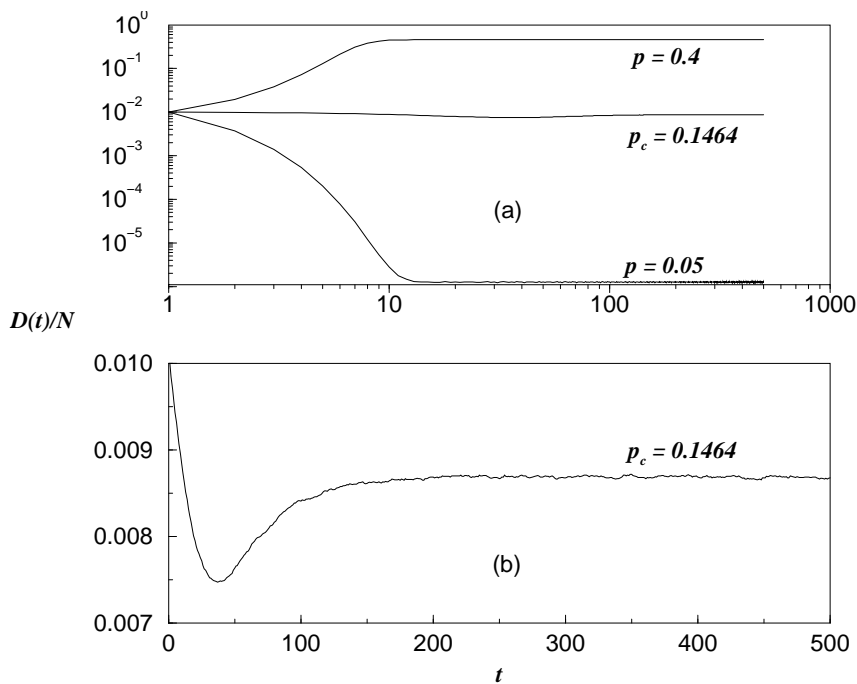,width=\hsize,clip=}
\centering
\caption[]{ 
  \small Hamming distance as a function of time for a Kauffman net
  composed of $N=10000$ elements and connectivity $K=4$.  (a) Log-log
  graph showing the Hamming distance for the three different regimes
  of the system: frozen ($p=0.05$), critical ($p_c = (1-\sqrt{1/2})/2
  \simeq 0.1464$) and chaotic ($p=0.4$). In all the cases the initial
  Hamming distance was $D(0)=100$. (b) Hamming distance for the
  critical phase ($p=p_c$) but in a non-logarithmic graph. Note that
  the Hamming distance initially decreases, and then it rises again to
  saturate at a constant value that depends weakly on system size.}
\label{nkhamming}
\end{figure}
%
%
%
%
\begin{figure}[h]
\psfig{file=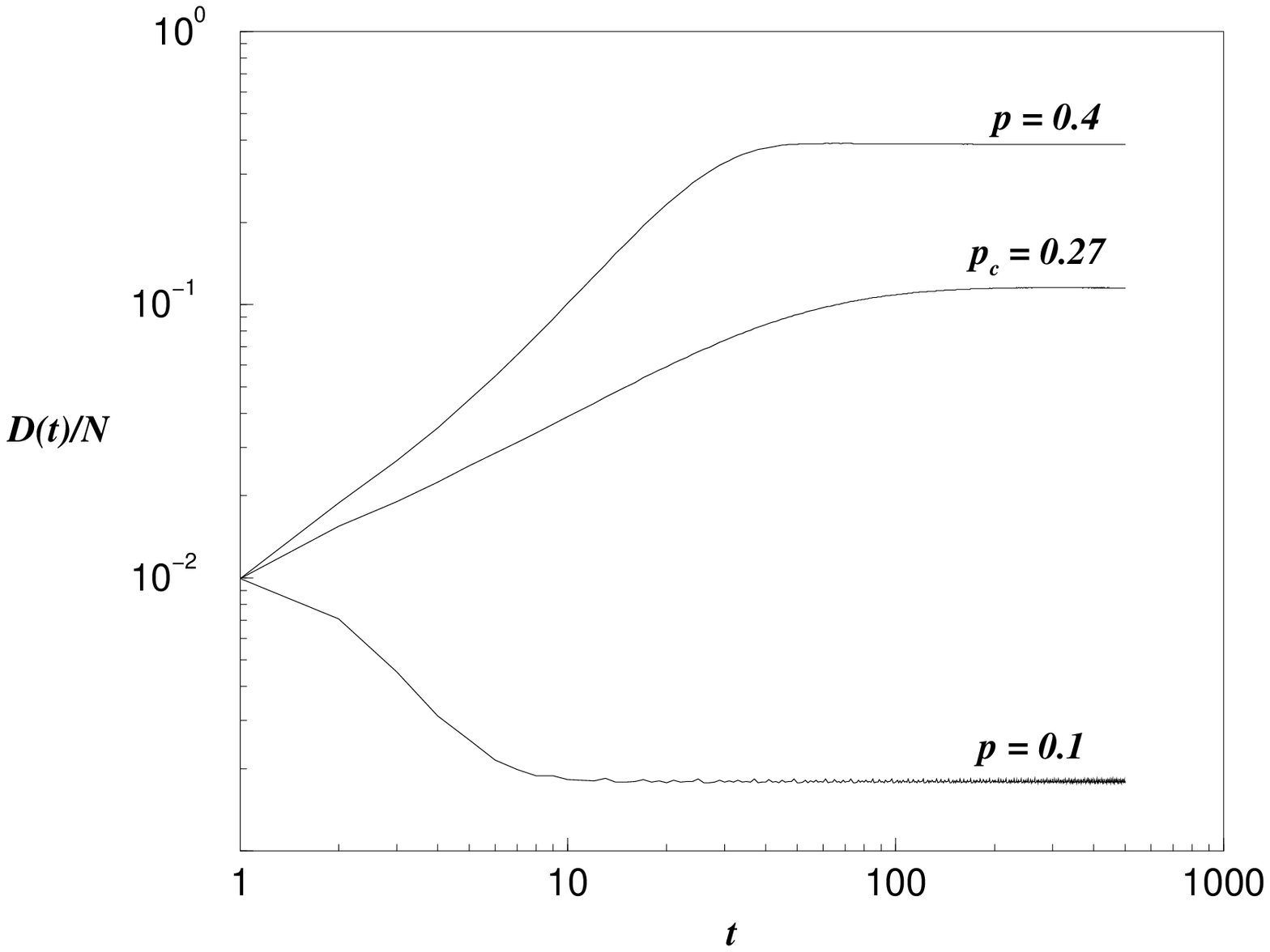,width=\hsize,clip=}
\centering
\caption[]{
  \small Hamming distance in a two-dimensional lattice composed of $N
  = 100\times 100$ elements. Every node in the lattice receive inputs
  from its four first nearest neighbors ($K=4$).  The three curves (in
  log-log scale) show the behavior of the Hamming distance in the
  three regimes: frozen ($p=0.1$), critical ($p_c=0.27$) and chaotic
  ($p=0.4$). Note that in the critical phase the Hamming distance
  initially increases algebraically and then saturates at a constant
  value that depends on system size.}
\label{lthamming}
\end{figure}
%
%
%
%

The behavior of the normalized Hamming distance, $D(t)/N$, can be seen
in figures \ref{nkhamming} and \ref{lthamming}, which respectively are
for the Kauffman net and a two-dimensional lattice system. In both
cases the system has $N=10^4$ elements and the connectivity is $K=4$.
The linkages of every element of the Kauffman net are chosen randomly,
whereas in the two-dimensional lattice each element receives inputs
from its four nearest neighbors. Both figures contain three curves,
with the parameter $p$ picked to put the systems into the three
different phases. For the Kauffman net $p_c = (1-\sqrt{1/2})/2$ (see
equation (\ref{K_c}) below). The value of $p_c$ is not very well known
for the two-dimensional lattice, but the best numerical estimations
indicate that $p_c\simeq0.29$ for an infinite lattice
(\cite{88.6,88.4,87.7,87.1,86.4}). For finite lattices the value of
$p_c$ has been defined as the average over realizations of the value
of $p$ at which a first cluster spanning the whole net appears
(\cite{88.8}). For a $100\times 100$ lattice this value is $p_c\simeq
0.27$.

In the frozen phase the distance has an initial transient but then
quite quickly approaches an asymptotic value.  In the chaotic phase
the distance shows an exponential rise followed by a similar
saturation. These behaviors are almost the same for both the Kauffman
net and the two-dimensional lattice.  On the other hand, in the
critical phase the behavior of the Hamming distance is very different
in these two systems.  In the Kauffman net the distance initially
decreases and then increases again, asymptotically approaching a
constant value that depends weakly on system size.  In contrast, for
the lattice the Hamming distance initially grows and then
saturates. We will see later that, within the framework of the
annealed approximation, the normalized Hamming distance for an
infinite Kauffman net approaches zero monotonically in both the frozen
and the critical phases (exponentially in the frozen phase, and
algebraically in the critical phase).  As far as we can tell, the
non-monotonic behavior of the Hamming distance in finite systems at
$K_c$ shown in Fig. \ref{nkhamming}b has not yet been explained.

\subsubsection{Limitations on the mean field calculation}

Let us look back at the argument which led to equation (\ref{expgrow}).
Calculations like this, in which actual system properties are replaced
by average system properties are in general called ``mean field"
calculations.

Naturally, the results derived as equation (\ref{expgrow}) depend
crucially upon the assumptions made.  The derivation follows from the
assumption that the $f_i$'s in each step are effectively random. (See
also the derivations of equations (\ref{over}) and (\ref{overp})
below, which also depend upon the randomness assumption.) The
randomness will certainly be correct in the annealed situation, in
which the couplings are reshuffled in each step.  It will also be true
in Kauffman net in the limiting situation in which $K=\infty$.  In
that case, information is spread out over the entire system and thus
has a very small chance of correlation among the $f_i$'s.  The
Kauffman net has a likely configuration that permits the replacement
of the actual values of the $f_i$'s by their statistical distribution
(\cite{87.3}). However, the approximations used here will not always
work.  Specifically, they fail in all kinds of finite $N$ situations,
or in situations in which the linkages are arranged in a
finite-dimensional lattice. In that case, the assumed randomness of
the $f_i$ does not hold, because their arguments are not random, and
the derived equations will not work.  To see this in the simplest
example choose $K=N=1$ with quenched couplings.  A brief calculation
shows that for any one of the four possible $f_i$'s, after a one-step
initial transient $a(t+2)=a(t)$.  That does not agree with equation
(\ref{over}) derived below.  In fact, for any finite dimension and
linkages which involve short-range couplings, the overlap is not unity
at long times even in the frozen phase.
More generally, if the system is put onto a finite dimensional
lattice, or if the functions are not picked at random, or if the
initial elements are not random, couplings initially used can be
correlated with couplings used later on. Then the information transfer
will be different and equation (\ref{expgrow}) will fail.

However, the principle that there can be a phase transition in the
kind of information flow remains quite true for a $d$-dimensional
lattice, and for other ensembles of coupling functions.

\subsubsection{Connections to percolation problems}

The transfer of information just described is quite similar to the
transfer which occurs in a typical phase transition problem.
Generically, these problems have three phases: ordered, critical, and
disordered (\cite{76.1,00.9}).  The bare bones of such an information
transfer problem is described as a {\em percolation problem}
(\cite{85.3}).

In one sort of percolation problem one considers a lattice.  Each bond
or link of the lattice is picked to be either connected or
disconnected.  The choice is made at random and the probability of
connecting a given link is picked to be $q$. Now one asks about the
size of the structure obtained by considering sets of links all
connected with one another.  For small values of $q$, these sets tend
to be small and isolated.  As $q$ is increased, the clusters tend to
get larger. At sufficiently large values of $q$, one or more connected
clusters may span the entire lattice.  There is a critical value of
the probability, denoted as $q_c$, at which such a spanning cluster
just begins to form. Properties of the large clusters formed near that
critical point have been studied extensively (\cite{85.3}).  The
resulting behavior is ``universal'' in that for an isotropic
situation, the critical properties depend only upon the dimensionality
$d$ of the lattice, at least when $d$ is less than four.  For $d>4$,
the percolating system obeys mean field equations.  When the symmetry
properties change, the critical behavior does change. For example, a
system with directed percolation (\cite{85.3,97.11}) has the
information flow move preferentially in a particular direction. The
critical behavior of directed percolation is different from that of
ordinary percolation.  It is attractive to explore the connection
between the phase transition for percolation, and the one for the
$N$-$K$ model.

Several authors have constructed this connection in detail.  For
example, \cite{88.3} looked at the $N$-$K$ model on lattices for $2,3$
and $4$ dimensions and demonstrated numerically that the phase
transition occurred when the information transition probability
reached the critical value for percolation on the corresponding
lattice.  \cite{88.4} showed the connection to percolation for both
sequential and parallel updating for a two dimensional lattice.
However, \cite{88.5} took very special forms of the connections, using
only rules with connectivities of the form of a logical ``or".  This
did not result in behavior like simple percolation but instead quite a
different phase transition problem, related to the behavior of diodes
and resistors.  At roughly the same time, \cite{88.6} indicated a
close numerical correspondence to the two dimensional percolation
problem both in $p_c$ and also in the fractal dimension of the
spanning cluster.  (For $p_c$ see also \cite{88.8}.) He also defined
an order parameter, essentially a Hamming distance, that, when plotted
as a function of $(p-p_c)$, looked just like a typical critical
phenomena result. He argued that the $N$-$K$ model is equivalent to
directed percolation.  More specifically, \cite{89.7} argued that the
quenched problem has a critical point which is in the same
``universality class'' (\cite{00.9}) as directed percolation.  This
would imply that the critical point is essentially the same as that of
the directed percolation problem.  The qualitative properties of both
the ordered and the frozen phases would also be essentially similar in
the percolation case and the $N$-$K$ lattice. In the latter situation,
the preferred direction would be that of the ``time'' axis. The
structure of both would vary with dimensionality and become like that
of mean field theory above four dimensions.  This is the same mean
field theory which describes the Kauffman net.  Thus, the behavior of
information transfer in $N$-$K$ problems was mostly understood in
terms of percolation.

\subsection{Lattice versus Kauffman net}

We can now point to an important difference between systems in which
all elements are coupled to all others, as in the Kauffman net, and
lattice systems in which the elements which are ``close" to one
another are likely to be mutually coupled.  ``Closeness'' is a
reciprocal relation.  If $a$ is close to $b$, then $b$ is also close
to $a$.  Therefore, close elements are likely to be coupled to one
another and thereby form a closed {\em linkage loop}. Any large-$N$
lattice system might be expected to form many such loops.  When $K$ is
small, different spatial regions tend to be unconnected and so many
different modules will form\footnote{A module is a loop or more
complex topology of dependencies in which all the functions are
non-constant, plus all other elements that are influenced by that
structure. See section \ref{sec:loops}.}.  The dynamics of the
elements in different modules are independent.  In contrast, in a
Kauffman net, influence is not a reciprocal relation. If element
$\sigma_j$ appears in the coupling function $f_i$ associated with
element $\sigma_i$, there is only a small chance, proportional to
($K/N$), that $\sigma_i$ will appear in $f_j$.  For large $N$ and
small $K$, the probability that a given element will participate in a
linkage loop will be quite small, so there will then be a small number
of modules.  When $K$ is small, the number of modules in uniformly
coupled systems grows more slowly than the system size, while in
lattice systems the number of modules is proportional to the size of
the system.  This distinction will make a large difference in the
cycle structure.\footnote{The interested reader will recall that in
quantum field theory and statistical mechanics, mean field theory
appears in a situation in which fluctuations are relatively
unimportant.  This will arise when the analog of linkage loops make a
small contribution.  Then, the physicist puts the effect of loops back
into the problem by doing what is called a \emph{loop expansion}.
Since they both expand in loops, the percolation mean field theory and
the mean field theory of the Kauffman net are essentially the same.} 
For the flow of information the difference between the two kinds of
nets is more quantitative than qualitative.  One can see the
similarity between them by comparing the curves shown in figures
\ref{nkhamming} and \ref{lthamming}.
 
\subsection{Calculations of overlap and divergence} 
\label{percolateO}

Before coming to a careful description of the phases, we should
describe more fully the kinds of calculation of overlap that can be
performed. Equation (\ref{expgrow}) is just the beginning of what can
be done with the trajectories of states in this problem. In fact,
exactly the same logic which leads to that equation can give a much
more general result.  If the overlap between two states at time $t$ is
$a(t)$, and if the elements which are different arise at random, then
the probability that the arguments of the function $f_i$ will be the
same for the two configurations is
\begin{equation} 
\rho=[a(t)]^K .
\label{qtemp} 
\end{equation} 
If all arguments are the same, then the contribution to the overlap at
time $t+1$ is $1/N$. (The $N$ arises from the normalization of the
overlap.)  If one or more arguments of the coupling function are
different in the two configurations, and the functions $f_i$ are
picked at random, then the chance of having the same functional output
is $1/2$ and the contribution to the overlap is $1/(2N)$.  Since there
are $N$ of such contributions, weighted by $\rho$ and $1- \rho$
respectively, the equation for the overlap defined by equation
(\ref{overdef}) is
\begin{equation}
a(t+1)=\big[1+[a(t)]^K\big]/2 .
\label{over}
\end{equation}
There are several possible variants of this equation.  For example, if
the different outcomes 0 and 1 of the function $f_i$ are weighted with
probabilities $p$ and $1-p$ respectively, to produce a sort of
magnetization bias, then equation (\ref{over}) is replaced by
(\cite{86.3,86.4})
\begin{equation} 
a(t+1)=1 - \big[1-[a(t)]^K\big]/K_c ,
\label{overp}
\end{equation}
where $K_c$ is given in terms of $p$ as
\begin{equation} 
K_c=1/[2p(1- p)] .
\label{K_c} 
\end{equation}

%
%
%
%
\begin{figure}[h]
\psfig{file=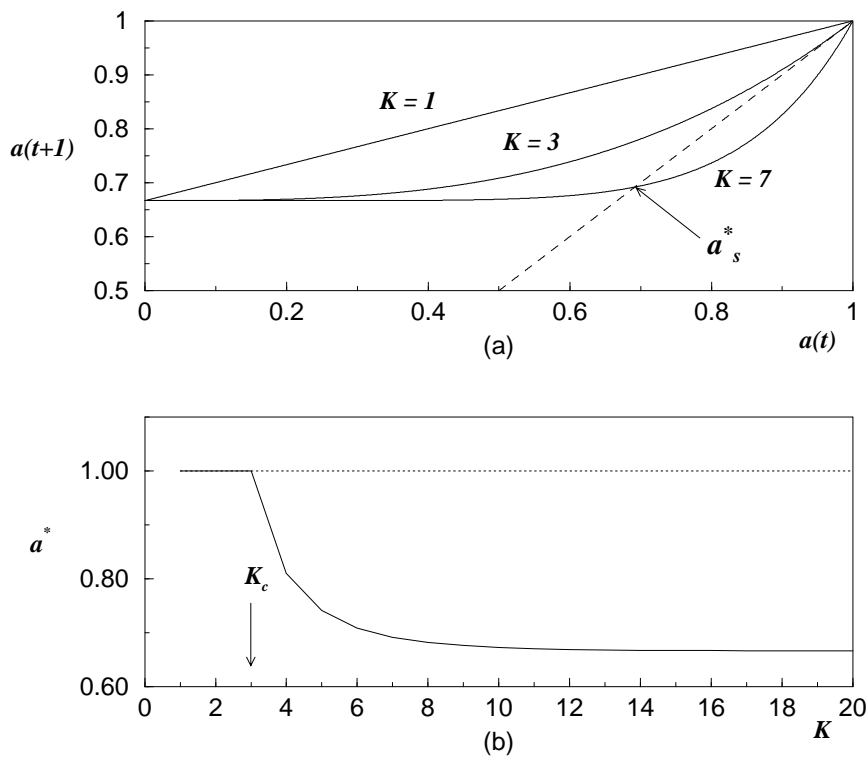,clip=,height=4in}
\centering
\caption[]{ \small (a) The mapping $F(a)=1-\left[1-a^K\right]/K_c$
(see Eq.~(\protect{\ref{overp}})) for $K_c = 3$ and three different
values of $K$ (solid curves), corresponding to the three different
phases of the system. The dotted line is the identity mapping. (b)
Bifurcation diagram of equation (\ref{over*}). For $K\leq K_c$ the
only fixed point is $a^*=1$. For $K>K_c$ the previous fixed point
becomes unstable and another stable fixed point $a^*_s$ appears.}
\label{mapping}
\end{figure}
%
%
%
%

In the limit $t\rightarrow\infty$, $a(t)$ asymptotically approaches
the fixed point $a^*$, which obeys, from equation (\ref{overp})
\begin{equation} 
a^*=1 -\big[1-[a^*]^K\big]/K_c .
\label{over*}
\end{equation}
We might expect equation (\ref{over*}) to reflect the three-phase
structure of the problem, and indeed it does. Fig.
\ref{mapping}a shows the graph of the mapping 
$F(a)=1-\big[1-a^K\big]/K_c$ for different values of $K$, and
Fig. \ref{mapping}b shows the bifurcation diagram of equation
(\ref{over*}). Both graphs were calculated with $p$ chosen so that
$K_c=3$.  As can be seen, if $K\leq K_c$ there is only one fixed point
$a^* = 1$, whereas for $K>K_c$ the fixed point $a^*=1$ becomes
unstable as another stable fixed point, $a^*_s\neq 1$, appears. The
value of the infinite time overlap $a^*$ describes the fraction of
elements whose value is sensitive to the cycle entered.

When $K>K_c$, the system is chaotic; $a^*$ is less than one even when
the starting points are quite close to one another. This reflects the
fact that the system has a long-term behavior which can include some
cyclic oscillations, the initial data determining the cycle entered.
(We discuss the cycles in detail in the next chapter.)  As $K$
approaches $K_c$ from above, $a^*$ increases since fewer elements have
final values which are sensitive to the initial data. On the other
hand, for $K\leq K_c$, the infinite time overlap is exactly one and
therefore the proportion of elements whose final value depends upon
the starting point is precisely zero.  Thus, independently of the
starting point, the system is always stuck in essentially the same
configuration.  This surprising result pertains to the Kauffman net.
In contrast, for any lattice system with $K> 0$ the final overlap is
less than unity, reflecting the fact that the system can store
information about initial data in some finite proportion of its
elements. Such storage is impossible for the Kauffman net.

\begin{figure}[h]
\centering
\psfig{file=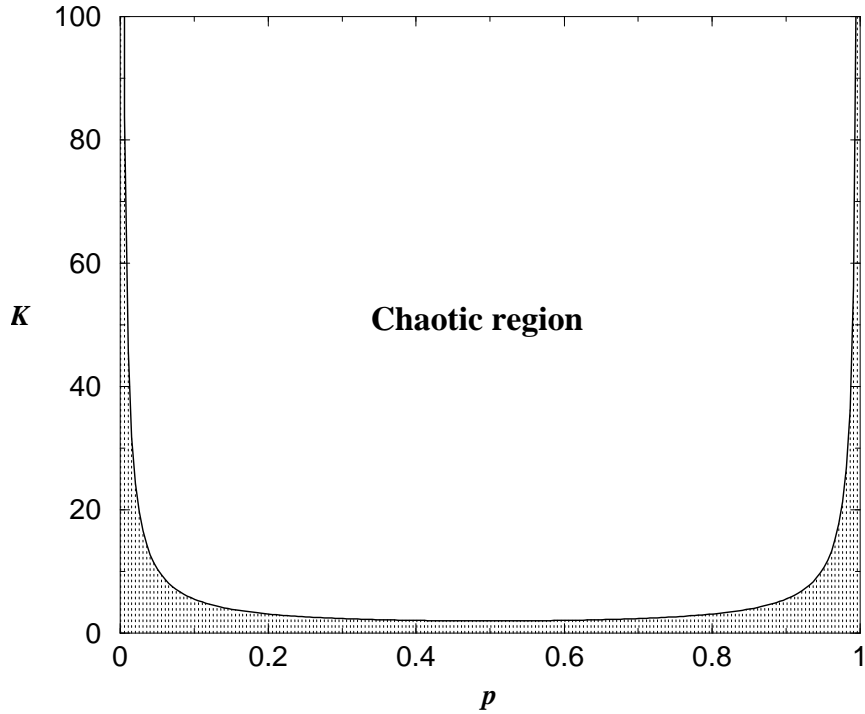,width=\hsize,clip=}
\caption[]{ \small Phase diagram for the $N$-$K$ model. The shaded
  area corresponds to the frozen phase, whereas the upper region
  corresponds to the chaotic phase. The curve separating both regions
  is the critical phase $K_c=[2p(1-p)]^{-1}$.}
\label{critical_line}
\end{figure}

It is worth noticing that for a given value of $K$, the system can be
put in one of the three different phases by adjusting the value of the
probability bias $p$. The critical value of $p$ is then obtained by
inverting equation (\ref{K_c}), which gives the \emph{critical line}
separating the frozen phase from the chaotic phase in the $p$--$K$
parameter space, as illustrated in Fig. \ref{critical_line}.

%
%
%
%
%
%
%
%

The rigidity of the Kauffman net was emphasized by Flyvbjerg
(\cite{88.10,89.6}), who wished to establish the equation for the
\emph{stable core}, the set of variables whose values at time infinity
do not depend upon initial data (see also
\cite{95.5,93.2,93.3,98.6}.)  He calculated the time 
dependence of the proportion of variables which had already settled
down, and thereby found a closed form for the size of the core. He
found that in the ordered state of the Kauffman net the
fraction of elements in the core is unity.
Another
kind of rigidity is studied by counting the \emph{weak elements}.  In a
particular realization, elements are termed weak if changing their
value never affects the long-term behavior of the system. Many authors
have studied these elements (see for example \cite{95.5,93.2,93.3}).
In the Kauffman net, but not in the lattice system, the proportion of
weak elements is unity throughout the frozen phase.

\section{Cycle Behavior}
\label{CycleBehavior}

\begin{figure}[h]
\centering
\psfig{file=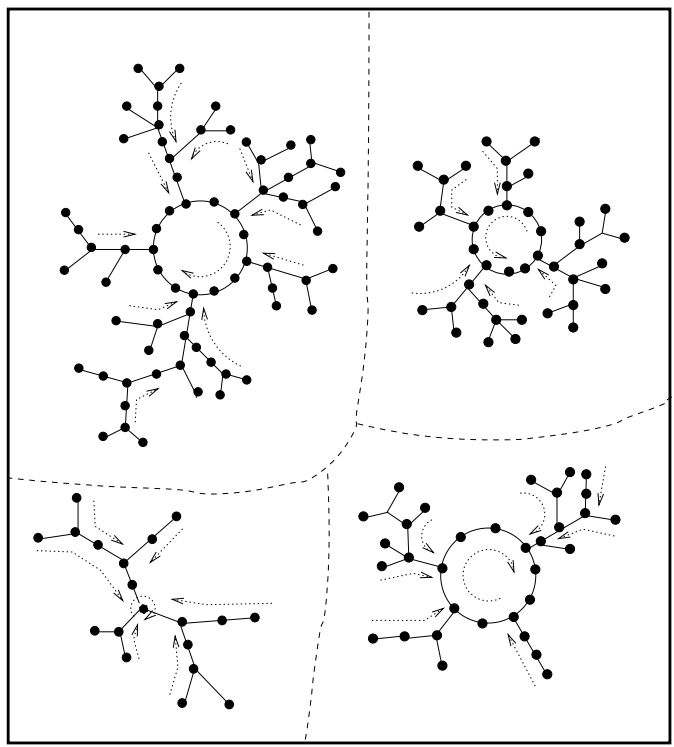,width=3.5in,clip=}
\caption[] { \small Schematic representation of the state space
of the $N$-$K$ model.  Each state is represented as a bold point.
Under the quenched dynamics, the state space is broken down into
several cycles, or attractors, represented as circles.  Each initial
state eventually will end up in one of these cycles (the arrows show
the direction of the flow).  The totality of states which evolve
towards a given cycle, is the basin of attraction of that cycle.  Note
that there can be attractors consisting of only one point,
corresponding to a cycle of period 1.}
\label{basins}

\end{figure}

For quenched systems the evolution functions $f_i$ are independent of
time. For any initial condition, the system eventually returns to a
previously visited point and then cycles repeatedly. The state space,
which consists of the $ \Omega = 2^N$ configurations of the system,
breaks up into different cycles and their basins of attraction, as
shown schematically in Fig. \ref{basins}.  Each initial condition will
eventually fall in one of these cycles, which are determined by the
evolution functions and the linkages.

%
%
%
The description of a cycle is, in some sense, much more complex than
the description of orbit separation.  In separation, one is dealing
with a very gross property of the system: Different parts of it can
behave independently and additively in the separation process.  We
utilized this fact in our calculations of overlaps and Hamming
distances carried out in sections \ref{percolateI} and
\ref{percolateO} above.  On the other hand, to get a cycle to close,
each and every element must simultaneously return to its value at a
previous time-step.  This is a much more delicate process and makes
for much more delicate calculations.  As we saw, information flow in
$N$-$K$ systems is closely analogous with the well-studied behavior of
percolation problems.  In contrast, the behavior of cycles in $N$-$K$
models is something special, not closely analogous to any of the usual
calculations for the usual systems of statistical mechanics.  When the
$N$-$K$ system forms cycles, one can ask many quantitative questions.
One might wish to know the number of steps the system takes before it
falls into a cycle (called the \emph{transient} time), and about the
length of the cycle.  For each cycle, there is a basin of attraction,
which is the set of initial states which eventually fall into that
cycle.  Therefore, one can develop a quantitative description of the
distribution of cycle lengths, transient lengths, basin sizes,
etc. either within a given realization or averaged over realizations.

\subsection {Linkage Loops and Cycles}
\label{sec:loops}

\begin{figure} [h]
\centering
\psfig {file=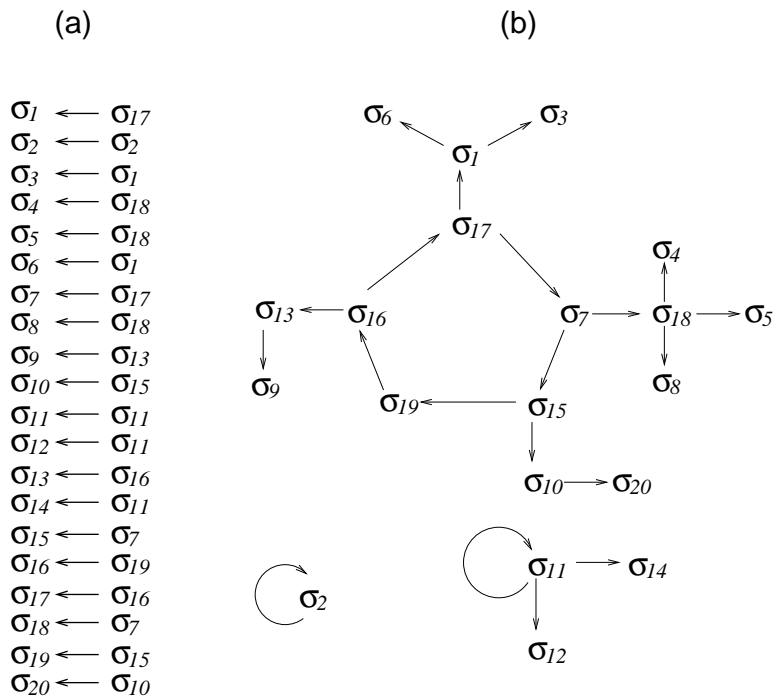, width=4in, clip=}
\caption[] { \small Linkage loops for an
$N$-$K$ net of $N = 20 $ elements and connectivity $K = 1 $.  (a)
Particular realization of linkage assignment in the net.  The first
column is the list of the $N$ elements of the net, $ \{ \sigma_1$, $
\sigma_2, \dots$, $ \sigma_N \} $, and the second column shows the
particular linkage every element has been assigned.  (b) Schematic
representation of the linkage loops.  Each arrow points towards the
descendants of a given element. In this particular realization there
are three modules, one of which consits of only one element,
$\sigma_2$.}
\label{linkageloops}
\end {figure}

The structure of the linkages is an important ingredient in
determining the nature of the cycles in the system.  Each element can
appear in the coupling functions for other elements.  These in turn
can appear in the couplings of others.  For each element one can trace
out its \emph{descendants}, i.e., the elements it affects.  Similarly,
one can chain backward and find the \emph {ancestors} for each
element, i.e., the elements which affect it.  For a cycle of length
longer than one to exist, at least one element must be its own
ancestor, and thus its own descendant.  (If no such element existed,
one could trace back and find elements with no ancestors.  They would
then have fixed values and be stable elements.  The elements
controlled by only them would also be stable.  The line of argument
would go forward until it was found that all elements were stable.)
Fig. \ref{linkageloops} illustrates the idea of ancestors and
descendants for the case $K = 1 $.  As can be seen, the connections
between different elements of the network give rise to
\emph{linkage loops} and \emph{linkage trees},
each tree being rooted in a loop.
%
\cite {88.1} stress the importance of
elements which are their own ancestors by pointing out that any
unstable element must be influenced by at least one such element.  Of
particular relevance are those elements which belong to a loop in
which there are only non-constant functions.  These self-influencing
elements, and the unstable elements that they influence, are called
the \emph{relevant elements} of the system.  They naturally form
themselves into groups called \emph{modules} (\cite { 99.8, 98.7,
98.10}).  Different modules do not influence one another and fully
determine the cycle structure.  Notice that the chain of linkages by
which an element can influence itself, namely its linkage loop, is
completely defined by the linkages, i.e. by the specific assignment of
the different elements in the $f_i$'s.  Linkage loops do not describe
the functions themselves.  Consequently, such loops are necessary but
not sufficient for the existence of non-trivial cycles.  Only loops
formed by relevant elements (with non-constant functions) are important
in determining the properties of the limit cycles of the network.  The
number of such cycles and their lengths depend crucially on the
modular organization of the relevant variables (\cite { 88.1, 98.6,
98.7}).  Consequently, linkage loops and relevant elements will form
an important part of our further discussions.

\subsection{Phase Transitions for Cycles.}

In the previous chapter, we saw that $N$-$K$ systems fall into
different phases, depending upon how effectively they transfer
information.  The cycles can be quite different in the different
phases.  The chaotic phase is characterized by very long cycles, in
which the typical cycle-lengths grow as a power of the size, $ \Omega
= 2^N$, of the entire state space.  Each cycle includes the
participation of many different elements which undergo a strongly
correlated dynamics.  In this phase, the transients are similarly long
and complex.  In contrast, the frozen phase tends to have much shorter
and simpler cycles, and also shorter transients in which the
individual modules do not grow as the system gets larger.  If the
system sits on a lattice of low dimension, the different modules are
spatially localized.  Each module shows a strong correlation in its
dynamics, but different modules move independently.  The critical
phase has larger modules and has a behavior intermediate between the
frozen and the chaotic phase.  These behaviors are known in
considerable detail.

The remainder of this chapter describes in detail the behaviors of the
different phases.  We start by discussing the exactly solvable models
which give a solid description of the limiting behavior of the phases.
We then describe how information obtained from simulations and the
exact solutions can be pieced together to give a qualitative
description of the phases.

\subsection{Soluble Models.}

\subsubsection{ Independent Subsystems}
\label{independ}

Here we do a preliminary calculation which will be of use in
interpreting results involving cycles.  Imagine a system composed of
$N$ independent subsystems.  Each subsystem has a probability $
\rho_l$ of having a cycle of length $l$.  We imagine that $ \rho_l$
gets quite small for large $l$ and ask what is the chance of finding a
long cycle in the entire system.  Notice that the chance of not
finding a piece with a cycle of length $l$ in the entire system is
\[q_l = (1-\rho_l)^N \approx \exp (-N \rho_l). \] If then, $ \rho_l$
varies exponentially with $l$ (the justification for this assumption
will be given below), namely, if
\begin{equation}
\rho_l = A \exp (-\alpha l),
\label{exponential_l}
\end{equation}
then we might expect to find parts with all cycle lengths up to
\begin{equation}
l_{mx} = (\ln N) / \alpha
\label{lmax}
\end {equation}
(so that $q_l$ is not that small, say of order $q_l \sim e^{-A}$).  To
make a long cycle in the entire system, one puts together many
sub-cycles of different lengths, $l_i, i = 1, 2, \dots, N$.  The total
cycle length, $L$, is the smallest number divisible by each of the
$l_i$'s.  Then $L$ will be a product of all prime numbers, $p_r$,
which are less than $l_{mx}$, each raised to a power $s_r$ which is
the largest integer for which the inequality
\[
\left[\frac {
l_{ mx} } { p_r^{ s_r} } \right] \geq 1 
\]
is satisfied ($[x] $ being the integer part of $x$).  Hence, to a
decent approximation, the largest cycle length $L_{mx}$ will be
\[L_{ mx} \approx (l_{ mx})^{ \pi (l_{ mx})} , \]
where $ \pi (l) $ is the number of primes less than $l$, which can be
estimated in the asymptotic limit of large $l$ as
\footnote {A better approximation is $ \pi (l) = 1 /(\ln l - 1)$
(\cite{62.1}).
Also, in this reference it is shown that
$l / \ln l < \pi (l) < 1.0423 l / \ln l$ for all $l > 10 $.}
\[
\pi (l) = l / \ln l. 
\]
In the end then, the longest cycle length $L_{ mx}$
obeys
\[\ln L_{ mx} \approx \pi (l_{ mx})(\ln l_{ mx}) \approx l_{ mx}, \]
so that
\begin {equation}
L_{ mx} \approx N^{ 1 / \alpha} .
\end{equation}

We have reach the remarkable conclusion that even though the
probability of long cycles in each component of the system falls
exponentially, the typical maximum cycle length in the entire system
depends algebraically upon the size of the system.  This calculation
does not apply directly to Kauffman nets because we have not accounted
for the fact that the different modules have different distributions
$\rho_l$, but nonetheless it is instructive.
\subsubsection{$K = 0$}
The case in which $K$ is zero is simple and uninteresting.  After the
first step, each element has a value which is completely determined by
its function $f_i$.  Each element remains with the value it had at
time one for all subsequent times.  Thus the system is completely
frozen.

\subsubsection{$K = 1 $}
In an important paper, \cite{88.1} analyze the structure of the case
in which each coupling function depends upon the value of just one
element.  To analyze this case the authors focus upon those spins
which are ancestors of themselves.  As we have seen (see
Fig. \ref{linkageloops}), each such element forms part of a coupling
loop of a length which we will denote as $m$.  No information may pass
into such a loop from other parts of the system.  Each element in the
loop may affect others, but the affected elements either are constant
or they inherit the cycle length of the coupling loop.  In a lattice
system, each element is coupled with a neighboring element.  In any
finite number of dimensions, there is a nonzero probability that two
nearest neighbor sites are inputs to each other.  Therefore, a system
with $N$ sites will contain a number of loops that is proportional to
$N$.  In contrast, on the Kauffman net, the couplings are not to
neighbors but randomly chosen from the whole system.  When $K = 1 $,
the probability that two sites are inputs to each other is
proportional to $1 / N$, and the average number of loops grows
logarithmically with $N$ (\cite { 98.7}).  Flyvbjerg and Kjaer
calculate the probability of observing $n_m$ loops of length $m$ in a
system of $N \rightarrow \infty$ elements.  Let $m_T$ be the total
number of elements contained in loops
\[m_T = \sum_{ m = 1} ^\infty m n_m.  \]
Then the distribution of $ \{ n_1, n_2, \dots \} $ takes the form
\begin{equation}
P (n_1, n_2, \dots) = \frac
{m_T} {N} \exp[-m_T^2 / (2 N)] \prod_{ m = 1}^{\infty}
	\frac { (1 / m^{n_m})} { n_m !} .
\label{m-distrib}
\end {equation}

Thus one can have reasonably long loops with a maximum likely length
of order $N^{ 1 / 2} $.  The linkage loops do not determine the cycle
structure.  To know the number and kind of cycles, one has to know the
coupling functions.  For a given loop to make a non-trivial cycle, all
the functions on it must be either identity or negation, for if there
is a constant function in the loop, one element has a fixed value and
it passes on its constancy to all the other elements in the loop.  So
think of a specific case.  Let us have a loop, with $m = 3 $, in which
element 1 is the ancestor of element 2 which is the ancestor of
element 3 which is the ancestor of element 1.  Let all the coupling
functions be identity.  Then the initial data just cycles around the
loop.  If the initial data is $ (ABC) $ then next is $ (CBA) $, etc.
There are two cycles of length one in which all elements are identical
and two of length three in which they are not.  On the other hand, if
all the coupling functions of this three elements are negation, there
is one cycle of length two and one of length six.  All the different
situations are similarly easy to analyze.
%
\begin{figure}[h]
\centering
\psfig{file=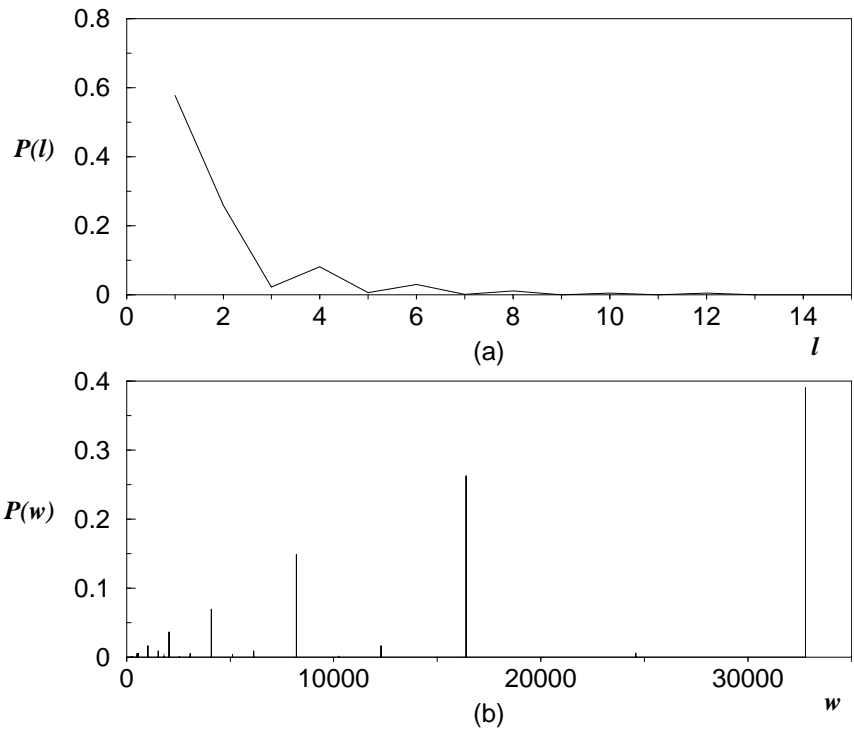, width=\hsize, clip =}
\caption[] {\small (a) Distribution of the probability that a cycle has length
$l$, $P (l) $, as a function of $l$ for a Kauffman net of $N = 15 $
elements and $K = 1 $.  Note that the probability of having cycles
with lengths larger than 10 is rather small.  (b) Probability $P(w) $
that an arbitrary state belongs to a basin of attraction of size $w$,
also for $N=15 $ and $K=1$.}
\label{cycles_k1}
\end {figure}

Despite the fact that the Kauffman net can contain very large linkage
loops for $K=1 $, the cycles are reasonably short.  Fig. \ref
{cycles_k1}a shows the distribution of cycle lengths in a Kauffman net
composed of $N=15 $ elements, and Fig. \ref {cycles_k1}b shows the
probability $W (n) $ that an arbitrary state belongs to a basin of
attraction of size $n$.  For $N=15 $, there are $2^{15}=32768 $
different states.  Nonetheless, as can be seen from the figure, the
probability of finding a cycle with length bigger than $10$ is
negligibly small.  The reason is that the probability that a loop is
relevant (has only non-constant functions) decreases exponentially as
the loop length increases.  In the usual calculation one assigns the
$K=1 $ functions with equal weight.  If either of the two constant
functions are present in the loop, the cycle length is unity.  If $a$
is the probability of assignment of the constant functions and $b=1 -
a$ is the probability of assignment of the other two functions,
identity and negation, the probability of finding a cycle of length
$l$ is proportional to $b^l$.  In most calculations $b$ is $1 / 2 $,
so that the probability of really long cycles falls off exponentially
in $l$ (this is the justification of equation (\ref{exponential_l})).
\cite{88.1} point out that in the special case with $b=1 $,
the probability of long cycles falls algebraically.  They speculate
that the behavior in this limit might be, first, analyzable, and
second, very similar to the behavior of the Kauffman net in the
critical case ($K=K_c$) described below.  As far as we know, these
speculations remain unproven.
\subsubsection { $K \geq N$}

Another case in which it is possible to analyze the structure of the
cycles in great detail is the one in which each coupling function
depends upon all the values of all the different elements in the
system.\footnote{The results of this section have been known for quite
some time; see references \cite{60.1,87.9}, which study what is called
the \emph{random map model}.  Here one studies systems in which one
has a random map from point to point in configuration space.  The form
of argumentation in this section closely follows
\cite{01.2}.}
This case has the simplifying characteristic that a change in a single
spin changes the input of every coupling function.  Therefore, one can
analyze some features of the behavior of the system as if the system
were annealed rather than quenched.  In particular, one can calculate
probabilities for hopping from configuration to configuration as if
the system were undergoing a random walk through a space of size $
\Omega$.  The quenched nature of the system only asserts itself when
the hopping takes onto a configuration previously visited.  After
that, one can be sure that the subsequent behavior will be cyclic.
There are almost classical mathematical analyses of this situation
(\cite{60.1,87.9}).  We describe this case by considering the
calculation of typical distributions of cycle lengths and of
transients.  Imagine a Kauffman net with $K \geq N$.  Imagine that we
start from a random configuration $\Sigma_0$ at time $t = 0 $.  At
subsequent times, we step by step follow the dynamics and essentially
go through a random walk $ \Sigma_0$, $ \Sigma_1$, $ \Sigma_2, \dots$,
through the configuration space, which has size $ \Omega$.  This walk
continues until we land upon a point previously visited.  Let us
define $q_t$ as the probability that the trajectory remains unclosed
after $t$ steps.  If the trajectory is still open at time $t$, we have
already visited $t + 1 $ different sites (including the sites $
\Sigma_0$ and $ \Sigma_t$).  Therefore, there are $t + 1 $ ways of
terminating the walk at the next time step and a relative probability
of termination $ \rho_t = (t + 1) / \Omega$.  The probability of still
having an open trajectory after $t + 1 $ steps is
\[q_{ t + 1} = q_t (1 - \rho_t) =
q_t \left (1 - \frac { t + 1} { \Omega} \right)
\mbox { with} \ q_0 = 1, \]
while the probability $p_{ t + 1} $ of terminating the excursion at
time $t + 1 $ is
\[p_{ t + 1} = \frac { t + 1} { \Omega} q_t.  \]
To obtain $P (L) $, the probability that a given starting point is in
the basin of attraction of a cycle of length $L$, we note that a
closure event at time $t$ yields with equal probability all cycle
lengths up to $t$.  Therefore,
\begin{equation}
P (L) = \sum_{ t = L}^{ \Omega} \frac{ p_t}{ t} ,
\label{P_L}
\end {equation}
which, in the limit of large $ \Omega$ can be approximated by
\begin {equation}
P (L) \approx \int_{ L}^{ \infty} \frac { 1} { \Omega}
e^{ -x (x - 1)/ ( 2 \Omega) } dx
\label {P_L_approx}
\end {equation}
It is also important to consider the probability $P (m, L) $ of
finding a cycle of length $L$ after having gone through a precursor of
length $m$, given by
\[P (m, L) = \frac { 1 } { \Omega } q_{ m + L - 1 } .  \]
(The factor $1 / \Omega$ comes from the fact that only one point, $
\Sigma_m$, of the state space can split the entire sequence $ \{
\Sigma_0$, $ \Sigma_1, \dots, $ $ \Sigma_m$, $ \Sigma_{m + 1} ,
\dots$, $ \Sigma_{m + L} \} $ of $m + L$ states into two pieces of
lengths $m$ and $L$ respectively.)  In the limit of large $ \Omega$,
the previous expression can be approximated by
\begin{equation}
P (m, L) \approx
\frac { \exp[-(m + L)^ 2/ (2 \Omega)]}{\Omega} .
\label{P_m_L}
\end {equation}

The most important characteristic of the results (\ref{P_L_approx})
and (\ref{P_m_L}) is that the typical cycle length and the typical
precursor length are each of order $ \Omega^{ 1 / 2} $.  Thus, a very
small fraction of the total configurations participate in each
transient or cycle, but nonetheless the cycles and the transients may
be very long.  There is another, and very nice, interpretation of the
results just calculated.  If the precursor length is zero, equation
(\ref {P_m_L}) gives the probability that our system will contain a
cycle element in a cycle of length $L$.  Since there are $ \Omega$
possible starting points, the average number of cycles of length $L$
in our system is
\begin {equation}
\langle N_c (L) \rangle = \frac { \exp[-L^2 / (2 \Omega)]} { L} .
\label {N_c}
\end{equation}
Here the $ \langle \cdots \rangle$ represents an average over
realizations.  An integration then gives us the information that the
average total number of cycles is proportional to $ \ln \Omega$, or
more precisely\footnote {Another result, often reported in the
literature (\cite { 96.2}, or table 1 in \cite{90.3}) is $ \langle N_c
\rangle = N / e$.  We do not know the justification for this, and
suspect that it is wrong.}
\begin {equation}
\langle N_c \rangle = \frac {\ln 2} { 2} N + \mathcal { O} (1).
\label {N_ctot}
\end {equation}

\begin {figure}[h]
\psfig{file=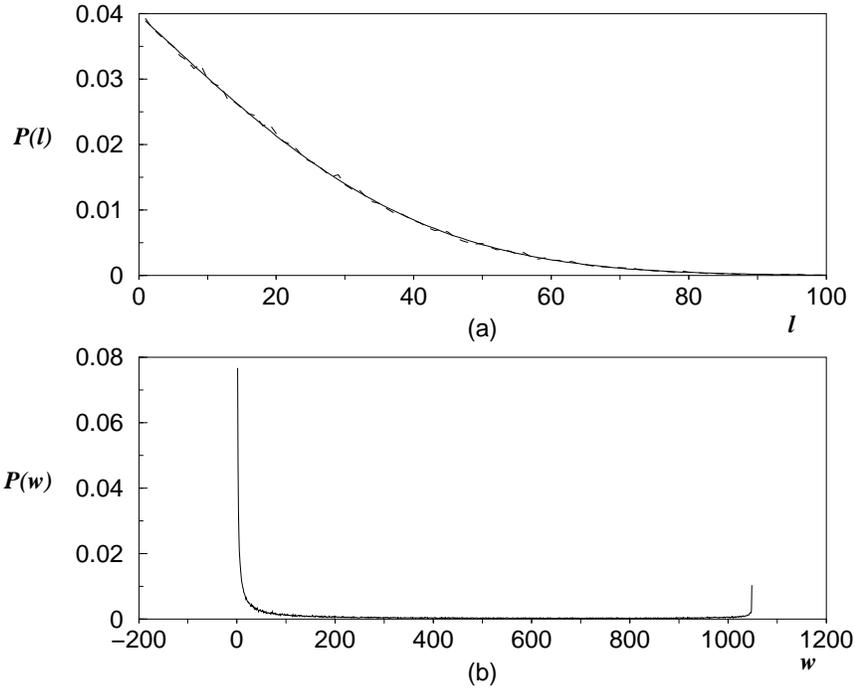, width=\hsize, clip=}
\centering
\caption{\small (a) Probability distribution of cycle lengths for
a Kauffman net with $N=K = 10 $.  The solid curve is the theoretical
result (\ref{P_L}) and the broken line is the result of our numerical
simulation.  (b) Probability that a given starting point $\Sigma_0$
belongs to a basin of attraction of size $w$, also for the case $N = K
= 10 $.}
\label{cycles_k10}
\end {figure}
%
%
Fig. \ref{cycles_k10}a shows the distribution $P (L) $ for a Kauffman
net with $N = K = 10 $, while Fig. \ref{cycles_k10}b shows the
probability $P(w)$ that a given starting point belongs to a a basin of
attraction of size $w$.  Note that most of the distribution $P (w) $
is concentrated around $w = 1 $ and $w = \Omega$, being nearly
constant (zero) in between.  The above reflects the fact that there
are large fluctuations in the number of cycles in different
realizations.  If the number of cycles were a typical extensive
quantity, its median would be the same as its average value.  Instead,
for large $N$ its median value is half the average, indicating that,
in the Kauffman net, the average is dominated by a few situations with
anomalously many cycles (\cite{60.1,87.9}).  The typical size of the
largest basin is given by dividing the entire space (of size $ \Omega
= 2^N$), by the number of cycles (of order $N$, which is actually of
order one, given the variance involved with the various fluctuations).
To form a basin of size $ \Omega = 2^N$, one starts from the cycle
with its typical size $2^{N/2}$ and count backward finding the set of
all first order ancestors, then backwards again to find their
ancestors, which are second order ancestors of the cycle, and so on.
Since these classes are non-overlapping and bounded in size, after
some number of steps (perhaps of order $2^{N/2}$, the typical
precursor size) the classes will shrink in size and eventually one
will terminate on one or a few of the most remote ancestors.  The
total basin size is the sum of the number of elements in these
ancestor classes.  Since each configuration may have no ancestors, one
ancestor or many ancestors, this backward-stepping, ancestor-counting
process will have a character similar to a multiplicative random walk.
As such it is an inherently highly fluctuating process.  So we might
expect huge fluctuations in the basin sizes (see
Fig. \ref{cycles_k10}b).
%
%
\subsection {Different Phases -- Different Cycles}
\subsubsection{Frozen Phases}
In the frozen phase, information typically propagates from a given
element to only a very few other elements.  Thus a change in initial
data will typically affect the subsequent behavior of only a few
elements.  Similarly, damage to a single coupling function will
produce changes which propagate to only a limited number of elements
in the system.  In these respects the lattice $N$-$K$ models and the
Kauffman net are very similar.  However, in other respects they are
very different.  For example, in the lattice system, the number of
relevant modules is proportional to $N$, so each initially different
element has a nonzero chance of being in a relevant module, and each
difference at time zero has a nonzero probability of developing into a
difference at infinite time.  In contrast, in the Kauffman net there
are typically only a few short cycles and the overlap rapidly
approaches unity, no matter what its initial value might have been.
For the $N$-$K$ model on the lattice there are many loops.  In one
dimension with nearest neighbor couplings all loops have length unity,
but with longer range couplings and in more dimensions some loops
might be quite long.  Despite the fact that at any given point in the
lattice it is exponentially improbable to find long loops, the
argument of section \ref{independ} indicates that the average loop
size might well vary algebraically with the size of the system.

\begin {enumerate}
\item {\bf Lattice system.} For small $K$, the lattice system falls 
into a phase in which there are many short cycles.  The number of
cycles grows as a power $ \mu$ of the volume of the state space, $
\Omega$; thus, the typical basin size will be $ \Omega^{ 1 - \mu} $.
The typical cycle length grows as a power of $N$, with multiplicative
fluctuations of order $ \ln N$.  The growth in cycle length occurs
because different modules will have sub-cycles of different lengths.
The entire cycle length is the product of the prime factors coming
from the modules.  Large primes are exponentially unlikely in a given
region, but the number of regions observed grows with $N$.

Compare the time development coming from two starting configurations
in which some small fraction of elements, spread out through the
system, are different.  These differences will each have a finite
probability of causing a different cyclical behavior, localized in its
particular region of the lattice.  Independent of the size of the
system, at long times the Hamming distance will go to a constant
$D^*$, proportional to $D (0) $ with a constant of proportionality of
order one.

\item {\bf Kauffman net.} Because the Kauffman net has
many fewer relevant modules than the lattice system, even for large
$N$, one will be able to observe realizations which always relax into
a time-independent behavior, a cycle of length one.  According to
\cite{98.6}, the average number of cycles observed in a given
realization will be independent of $N$.  Realizations with long cycles
will be exponentially unlikely.  However, when one gets a long cycle,
one can expect to have very many of them.  The frozen phase is one in
which almost all the variables end up with fixed values.  That is, for
a given realization, after a large number of steps most variables
approach a value which is independent of the initial state of the
system.  Thus the system behaves, for most of the variables, as if it
were highly frictional with the result that each variable comes to a
stop its own `best' value.  This is the reason that the overlap goes
to one as time goes to infinity in this phase.
\end {enumerate}

The majority of our knowledge of the Kauffman-net behavior of the
frozen phase comes from two calculations.
As discussed above, the Kauffman net with $K=1 $ was solved exactly by
Flyvbjerg and Kjaer (\cite {88.1}).  For the case $K=2 $, \cite{93.3}
has proven a group of theorems which apply on the frozen side of the
$K = 2 $ system.  To ensure the system was frozen, Lynch looked at the
case in which there is unequal weight to the four classes of $K = 2 $
functions, and demanded that the constant functions have a larger
weight than the non-canalizing ones.  This ensured that the Hamming
distance, as calculated by an approach like that in equation (\ref
{expgrow}), would decay exponentially.  He then proved that the system
was in a frozen phase by showing:
\begin {enumerate}
\item
{ Almost all elements in the system were in their stable state.}
\item
{ Almost all gates were weak, that is, changing their value does not
affect the cycle that is entered.}
\item
{ The length of the transient, i.e.  the number of steps before the
system enters its cycle, is bounded below by a constant times $ \ln
N$.}
\item
{ There is also a rather strong bound on the cycle length.  The bound
includes the statement that the median cycle-length is bounded by a
number which is independent of $N$.}
\end{enumerate}

\subsubsection {Chaotic Phases}

In contrast to the frozen phase, the chaotic phase is one in which a
nonzero fraction of the variables remain sensitive to initial
conditions.  In fact, most variables keep changing their values
throughout the time development of the system.  In the chaotic phase
the average length of limit cycles and of the transient which occurs
before the entry of the limit cycle both grow exponentially with $N$.
Because of this explosive growth as the system is made larger,
numerical investigations of orbits in the chaotic phase are limited to
small system sizes (see, e.g., \cite{98.6}).

As discussed above, the case when each input is coupled to all the
others can be solved exactly (\cite{86.3,87.9}), with the typical
cycle length growing with system size as $2^{N/2}$.  When $K$ is
finite but greater than the critical value $K_c$, both the Kauffman
net and the lattice models have typical cycle lengths that grow
exponentially with $N$.  This behavior reflects a complex network of
dependency loops in this regime.  Though it is plausible that the
details of the interconnected loop structures could be different for
lattice models and for the Kauffman net, they do not appear to lead to
marked differences in the behavior of the two types of models in the
chaotic regime.

\subsubsection{Critical Behavior}

We have already seen that there is a close connection between the
phase transition of the $N$-$K$ model and the standard percolation
transition of statistical physics. According to \cite{89.7},
random Boolean networks on nearest-neighbor d-dimensional
lattices belong to the `universality class' of directed percolation
with quenched disorder in d+1 dimensions. However, this
result is an argument, not a theorem.  The sharing of the universality
class means that the $N$-$K$ models share many of the detailed
properties of correlation and ordering with the directed percolation
models of standard statistical mechanics. More particularly, the
annealed $N$-$K$ model based upon nearest neighbor interactions is
equivalent to a directed percolation problem (\cite{86.4}).  Moreover,
when the interaction range is infinite (roughly corresponding to
infinite dimensions or a simplectic geometry), the quenched (or usual)
$N$-$K$ model is equivalent to the corresponding directed percolation
problem (\cite{86.3,86.1}).
 
%
%
%
%
\begin{figure}[h]
\psfig{file=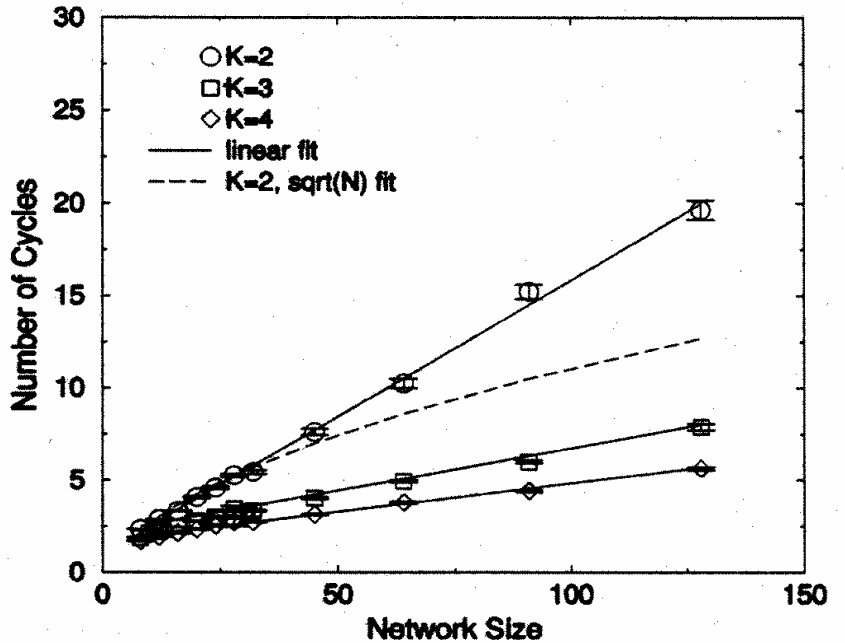,width=\hsize,clip=}
\centering
\caption{ \small Mean number of different limit cycles
as a function of the network size for critical Kauffman networks with
connectivities $K=2,3,4$. The solid curves are the best linear fits to
the numerical data, whereas the dashed curve is a $\sqrt{N}$ fit for
the case $K=2$. Taken from \cite{01.7} }
\label{fig:bilke}
\end{figure}
%
%
%
%

Many studies have been carried out to determine the statistical
properties of cycles in the critical phase. The main results can be
summarized by saying that in the critical phase, both the typical
cycle lengths as well as the mean number of different attractors grow
algebraically with $N$. More explicitly, it was believed that these
quantities were proportional to $\sqrt{N}$ (see for instance,
\cite{98.6,89.6,90.3,69.1,95.4,93.1}). This was one of the most
attractive results of the $N$-$K$ model in that it matched one of the
power-law behaviors exhibited by living organisms, as we will see in
section \ref{GenNet}.  However, S. Bilke and F. Sjunnesson have
recently brought into question the $\sqrt{N}$ dependence of the mean
number of different attractors, arguing that this result comes as a
consequence of a biased \emph{undersampling} of the whole state space
(\cite{01.7}).

Because the size of the state space grows exponentially with the
number of elements $N$, it is not feasible to enumerate completely all
the different cycles in a given network realization unless $N$ is
quite small.  What people usually do is to probe the state space with
a small fraction of its elements (which are assumed to be
representative), and then to infer the statistical properties of the
model from this sampling.  However, this method, which has yielded the
$\sqrt{N}$ dependence referred to above, has the problem that it is
possible to miss a small but still important fraction of cycles.

In their work, Bilke and Sjunnesson (ibid.) use a different
approach. They present a decimation method which eliminates the stable
elements of the network\footnote{A stable element is one which evolve
to the same fixed value independently of the initial state.},
retaining only the relevant ones. Since for any realization of the
network the number of different cycles as well as their lengths depend
only on the relevant elements, all the statistical information related
to limit cycles is preserved through this decimation procedure. The
advantage in eliminating the stable elements is that the number of
variables is drastically reduced. As a consequence, it is much easier
to perform a \emph{full enumeration} study of the number of different
cycles for each realization of the network. Through this approach,
Bilke and Sjunnesson found that, in the critical phase of the Kauffman
net, the mean number of different attractors $\langle N_c\rangle$
grows linearly with $N$ instead of as $\sqrt{N}$.
Fig. \ref{fig:bilke} (taken from \cite{01.7}) shows the mean number of
different cycles $\langle N_c\rangle$ as a function of the system size
$N$. As can be seen, the linear dependence $\langle N_c\rangle\sim N$
fits the numerical data much better than the $\sqrt{N}$ behavior.

In addition to the exponential growth of the state space as $N$
increases, the number of network realizations grows superexponentially
with $K$.  The evidence presented by Bilke and Sjunnesson raises the
possibility that some of the results obtained so far related to cycle
lengths and basins of attraction could also include a systematic bias
due to undersampling.  More work is needed to address these questions.

\section{Reversible Models}

Kauffman nets are generic models for $N$ elements coupling $K$
different variables.  In this section we specialize this generic model
to include an interesting symmetry property:
\emph {time reversal symmetry}.
The discrete symmetry and quenched
randomness together cause some new properties to emerge.

The standard Kauffman net that we have considered so far, whose
dynamics is given by equation (\ref {m1}), is dissipative because
multiple different states can map into one, so that information is
lost.  In this section we will refer to this system as the
\emph{dissipative Kauffman net} or simply as the \emph{dissipative N-K
model}.  Not all systems are dissipative; many of the systems
considered in Hamiltonian mechanics are \emph{reversible}.  Reversible
systems have the property that some transformation of the coordinates
(for example changing the sign of all velocities) makes the system
retrace perfectly its former path.  This section discusses some
aspects of the behavior of discrete reversible maps.  In the
time-reversible Boolean network studied in \cite{01.2, 01.3}, the
state of the system at time $t + 1 $ is governed by the equation
\begin{equation}
\sigma_i (t + 1) = f_i (\sigma_{ j_1 (i)} (t ),
\ldots, \sigma_{j_K (i)}(t)) \oplus \sigma_i (t - 1),
\label {treq}
\end {equation}
where the $\oplus$ denotes addition modulo $2$.  Each time-reversible
network realization has a corresponding dissipative realization with
the same functions and connections.  In the dissipative $N$-$K$ model,
the state $ \Sigma_{ t + 1} $ is completely determined by the previous
state $ \Sigma_t$.  Nevertheless, from equation (\ref {treq}) we see
that in the reversible model both $ \Sigma_{ t-1} $ and $ { \Sigma_t}
$ are required to calculate $ \Sigma_{ t + 1} $.  Thus, in the
reversible model the state of the system at time $t$ is represented as
\[S_t = { \Sigma_{ t-1} \choose \Sigma_t} , \]
and the state space has now $2^{ 2 N} $ points.  The behavior of the
reversible model is in some ways closely analogous and in other ways
quite different than that of the dissipative model.  Both models
exhibit a phase transition at $K_c$, a critical value of $K$ below
which the system ``breaks up'' into disconnected sections, and above
which there is a percolating cluster of connections.  The value of
$K_c$ is slightly lower for the reversible model $ (K_c \approx 1.6) $
than for the corresponding dissipative model ($K_c = 2 $), and some
details of the transition are different.  But for both models the
observed behavior is consistent with a percolation picture.  For $K <
K_c$, changing one element of a system leads to changes in only a
finite number of other elements, whereas for $K > K_c$, changing one
element causes a cascade of influence that spreads to a nonzero
fraction of all the other elements.  The typical cycle length grows
slower than linearly with the size of the system when $K < K_c$, and
exponentially with system size when $K > K_c$.

One big difference between the reversible and dissipative models is
that they have vastly different numbers of attractors (or, in the
reversible case, limit cycles).  Large differences in the behavior are
not entirely unexpected because in the dissipative model many
different state space points end up at the same attractor, whereas in
the reversible model every state space point is on exactly one limit
cycle.  For example, when $K = 0 $ (all input functions either $1 $ or
$0 $), the usual Kauffman net has only one attractor, while the number
of limit cycles of a reversible Kauffman net is proportional to $2^{ 2
N} $.  The reversible result can be understood by noting that for
either input function, each element is in one of four different
cycles, depending on its initial conditions, and that when $K = 0 $
the elements are all independent.  In the other limiting case $K = N$,
as discussed above, the number of attractors of a dissipative Kauffman
net is proportional to $N$ (see equation (\ref {N_ctot})).  In
contrast, in the reversible model with $K = N$, and indeed throughout
the regime $K > K_c$, the number of limit cycles grows as $2^N$.

The number of attractors in the reversible model with $K = N$ can be
understood by studying the mechanisms that lead to orbit closure.  One
way to close an orbit is to repeat two successive configurations, so
that $ \Sigma_T = \Sigma_0$ and $ \Sigma_{ T + 1} = \Sigma_1$.  Using
the approximation that each successive $ \Sigma$ is chosen randomly
from the $2^N$ possibilities yields a probability of an orbit closure
at a given time of order $ (2^{ -N})^2 = 2^{ -2 N} $.  However, there
is another mechanism for orbit closure that leads to much shorter
orbits and thus totally dominates the behavior.  If at some time $
\tau$ one has $ \Sigma_\tau = \Sigma_{ \tau-1} $, then
time-reversibility implies that $ \Sigma_{ \tau + n} = \Sigma_{
\tau-1-n} $ for all $n$.  Similarly, if $ \Sigma_ { \tau + 1} =
\Sigma_{ \tau -1} $, then $ \Sigma_{ \tau + n} = \Sigma_{ \tau-n} $
for all $n$.  Because the orbits reverse at these points, we call them
``mirrors.'' Once two mirrors have occurred, then the orbit must close
in a time that is less than the time it has taken to hit the second
mirror.  Again assuming that each successive $ \Sigma$ is chosen
randomly from the $2^N$ possibilities, one finds that a mirror occurs
at a given time with probability proportional to $2^{ -N} $, so the
expected number of steps needed to hit two mirrors is of order $2^N$.
Hence, typical orbit lengths are of order $2^{ N} $.  Since there are
$2^{ 2 N} $ points in the state space altogether, the number of limit
cycles is proportional to $2^N$.  In the reversible Kauffman net, when
$K$ is finite but greater than the critical value $K_c$, the
distribution of cycle lengths can be extremely broad.  For example,
when $N = 18 $ and $K = 2 $, the median cycle length $ \bar{l} $ is
approximately $ \bar {l} \approx 140 $, and yet the probability of
observing an orbit of length $10^{ 10} $ is greater than $10^{ -4} $.
This huge variability arises because of a nontrivial interplay between
the discrete symmetry and the quenched randomness in the system.  The
occasional extremely long orbit arises because some combinations of
coupling realizations and initial conditions are such that mirrors
cannot occur at all.  If the mirrors are not available to close the
orbits, then the system must wait until two successive configurations
happen to repeat.  In (\cite{01.2}) it is shown that in a system with
finite $K$ in the true thermodynamic limit, almost all realizations
and initial conditions yield no mirrors, and typical orbit lengths
grow as $2^{ 2 N (1-\epsilon)} $, where $ \epsilon$ is of order $1 /
(2^{ 2^K})$.  However, the crossover to the limiting behavior occurs
only when $N \sim 2^{ 2^K} $, so that even for moderate $K$, this
behavior is not accessible numerically (for example, when $K = 3 $,
one requires $N \sim 256 $, a value at which $2^N \sim 10^{ 77} $ and
$2^{ 2 N} \sim 10^{ 154} $).

The significance of this enormous variability in the cycle lengths is
not clear.  One way to interpret these results is to conclude that
characterizing cycles is not the right way to study the model.  It
would be interesting to investigate whether enormous variability
arises in other random systems with discrete symmetries, and to
determine whether there are possible experimental consequences.  One
possible starting point for comparison is with properties of random
magnets, for which regions of atypical couplings lead to Griffiths
singularities (\cite { 69.2}), which have stronger effects on dynamic
than static properties (\cite { 85.2}).  One must explore whether the
analogy is appropriate (this is not obvious because Kauffman networks
are not lattice-based), and if so, whether the results for the
reversible Kauffman model indicate that dynamics far from equilibrium
can be even more strongly affected by atypical coupling realizations
than are the properties in spin models that have been studied to date.

\pagebreak
\section {Beyond Cycles}
\subsection{Non-robustness of Cycles}

A cycle forms when the system returns to a configuration which is
exactly the same as one it has previously visited.  A demand for an
exact return should be viewed not as a single constraint but in fact
as $N$ constraints upon the configuration.  Such a strict demand makes
the properties of cycles quite special and delicate.  For this reason,
\emph{a study of cycles is probably not what one would want for
understanding the possible physical or biological consequences of the
models like the $N$-$K$ model.}  The statement just made flies in the
face of a very large body of work, some of which we have just
described.  We should, for this reason, argue for this statement with
some care.

Why don't we believe in cycles?
\begin {enumerate}
\item The characterizers of cycles are neither intensive
nor extensive variables.
\item With exponentially short
cycles in localized regions one gets power laws overall.
\item
The critical situation has very many, short cycles which are not
observed when one starts from randomly picked starting points.
\item
Attractor basins are complex in character, being at best multi-fractal
for large chaotic systems.
\item
In a large system you must wait so long to see a cycle that it cannot
be really important.
\item
In a large chaotic system, changing one rule changes the cycles quite
a bit.
\item
In both glasses and biological systems one wants to characterize the
system by things which are very robust.
\end {enumerate}

We do believe that generic properties of cycles are important though,
in that they characterize general aspects of dynamical systems.
Nonetheless, the huge fluctuations throughout realizations in such
important quantities as cycle lengths and number of different
attractors, calls for other types of characterizations.  Real
networks, whether they are genetic or neural or of any other kind, are
always subjected to external perturbations.  The robustness in the
dynamics of the network can not rely on quantities which change
dramatically with perturbations.  Hence, it is important to
characterize the dynamical properties of the network in the presence
of noise, trying to find out which kind of quantities are preserved
under the influence of noise, and which ones are not.

\subsection{Characterization via noise}

The addition of noise to a map provides a possibility for generating
additional information about the behavior of the models.  Noise
naturally blurs out the sharpness of behavior, making for a more
``fuzzy'' characterization.  Noise is, then, a natural way to get away
from the difficulty posed by the overly-precise characterization
provided by the cycles.

Unfortunately, most previous work on the $N$-$K$ model does not
include noise.  We do wish to point to two papers (\cite{89.3,89.5})
in which noise has been used to probe $N$-$K$ behavior.  For reasons
which will become more evident later on, we describe these papers
respectively as a ``crossing paper'' and a ``convergence paper". These
papers both break the precision of the dynamical rule of equation
(\ref{m1}) by saying that the rule is broken with a probability $r$
\begin{equation}
\sigma_i(t+1) =\left\{
\begin{array}{lcc}
f_i(\sigma_{j_1(i)}(t),\dots,\sigma_{j_{K_i}(i)}(t))
  &\mbox{with probability} & 1-r  \\
  \neg f_i(\sigma_{j_1(i)}(t),\dots,\sigma_{j_{K_i}(i)}(t))
& \mbox{with probability} &  r ,
\end{array}
\right. 
\label{m1cross}
\end{equation}
which can also be written in the alternative form:
\begin{equation}
\sigma_i(t+1)  = \left\{
\begin{array}{lcc}
f_i(\sigma_{j_1(i)}(t),\dots,\sigma_{j_{K_i}(i)}(t))
 & \mbox{with probability} & 1-2r  \\
  1  & \mbox{with probability}&  r   \\
  0  & \mbox{with probability}&  r .
\end{array}
\right.
\label{m1conv}
\end{equation}
This equation can be described as providing probabilities $r$ for
the two possible values of the outcome, independently of the value of
$f_i$.  In \cite{89.5} the rules are described in terms of a
temperature $T$, related to $r$ by 
\begin{equation}
r= \frac{1-\tanh (1/T)}{2} .
\label{Tdef}
\end{equation}

Both groups examine the development of two or more different initial
configurations using the same realizations.  They also apply exactly
the same rules and the same probabilistic choices to the different
configurations.  So far, both papers are essentially similar.  There
are two kinds of differences, the first being the choice of
measurement, and the second being the way they apply equations
(\ref{m1cross}) and (\ref{m1conv}).

The convergence paper starts with two or more randomly chosen initial
configurations, $\Sigma_0^{1}$, $\Sigma_0^{2},\dots$, $\Sigma_0^{m}$,
and calculates the resulting trajectories step by step:
\begin{eqnarray}
\Sigma_0^1 &\rightarrow& \Sigma_1^1\rightarrow\Sigma_2^1\rightarrow\dots
\Sigma_\tau^1 \nonumber \\
\Sigma_0^2 &\rightarrow& \Sigma_1^2\rightarrow\Sigma_2^2\rightarrow\dots
\Sigma_\tau^2 \nonumber\\
&\dots& \nonumber\\
\Sigma_0^m &\rightarrow& \Sigma_1^m\rightarrow\Sigma_2^m\rightarrow\dots
\Sigma_\tau^m \nonumber
\end{eqnarray}
At each step, and for each $i$, a choice is made among the three
branches of equation (\ref{m1conv}), and that choice is equally applied
to the $m$ configurations.\footnote{The two-branch versus three-branch
methods become inequivalent when they are applied to several
configurations at once.}  The calculation continues until two of the
$m$ configurations become identical (say for example
$\Sigma_\tau^1=\Sigma_{\tau}^2$).  The time needed to achieve the
convergence is noted.  We will denote this time by $\tau_m$,  stressing
the fact that $m$ configurations are being analyzed. 

In some ways, the convergence calculation is more complicated than the
crossing one.  The noise as defined by equation (\ref{m1conv}) tends to
produce convergence because it is applied equally to all trajectories
and because makes the values of the elements to be equal in all
trajectories.  In the limit of infinite $N$, the system shows three
phases: the low noise phase in which almost always trajectories will
not converge, the high noise phase in which trajectories will always
converge, and a separating critical phase.  These are respectively
described as low temperature, high temperature, and critical phases.

%
%
%
%
\begin{figure}[h]
\psfig{file=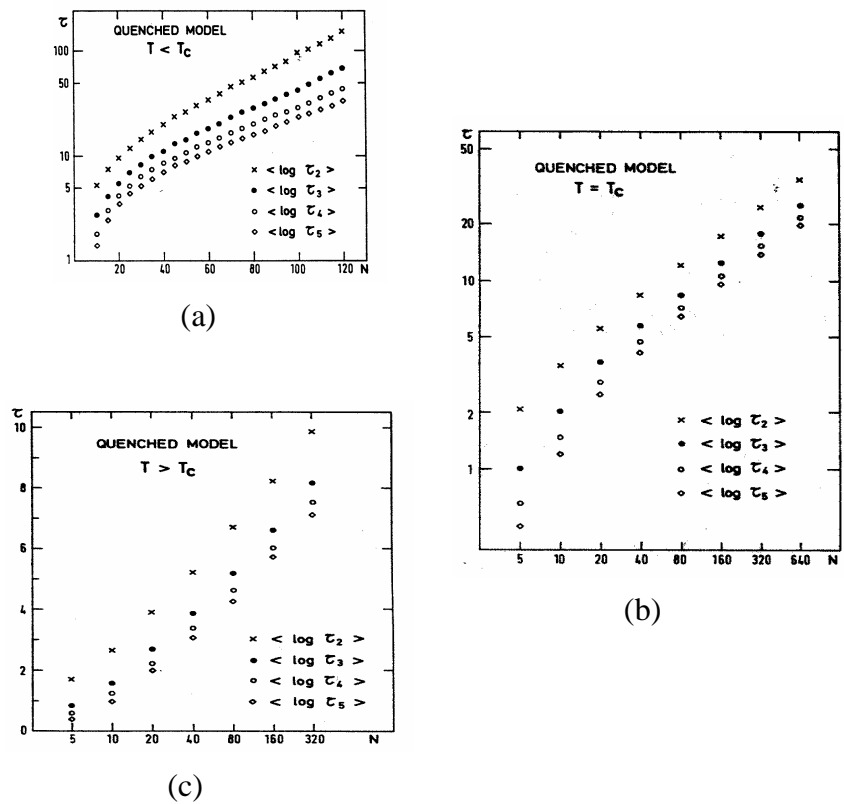,width=\hsize,clip=}
\centering
\caption{ \small Plots of $\langle\ln\tau_m\rangle$ versus $N$, taken 
from reference \cite{89.5}. In all the cases the connectivity of the
network was $K=4$. The three graphs correspond to three values of $r$:
(a) Low temperature phase, $r=0.15$; (b) Critical phase, $r=0.25$; (c)
High temperature phase, $r=0.35$.}
\label{Golinelli}
\end{figure}
%
%
%
%

%
%
%
%
\begin{figure}[h]
\psfig{file=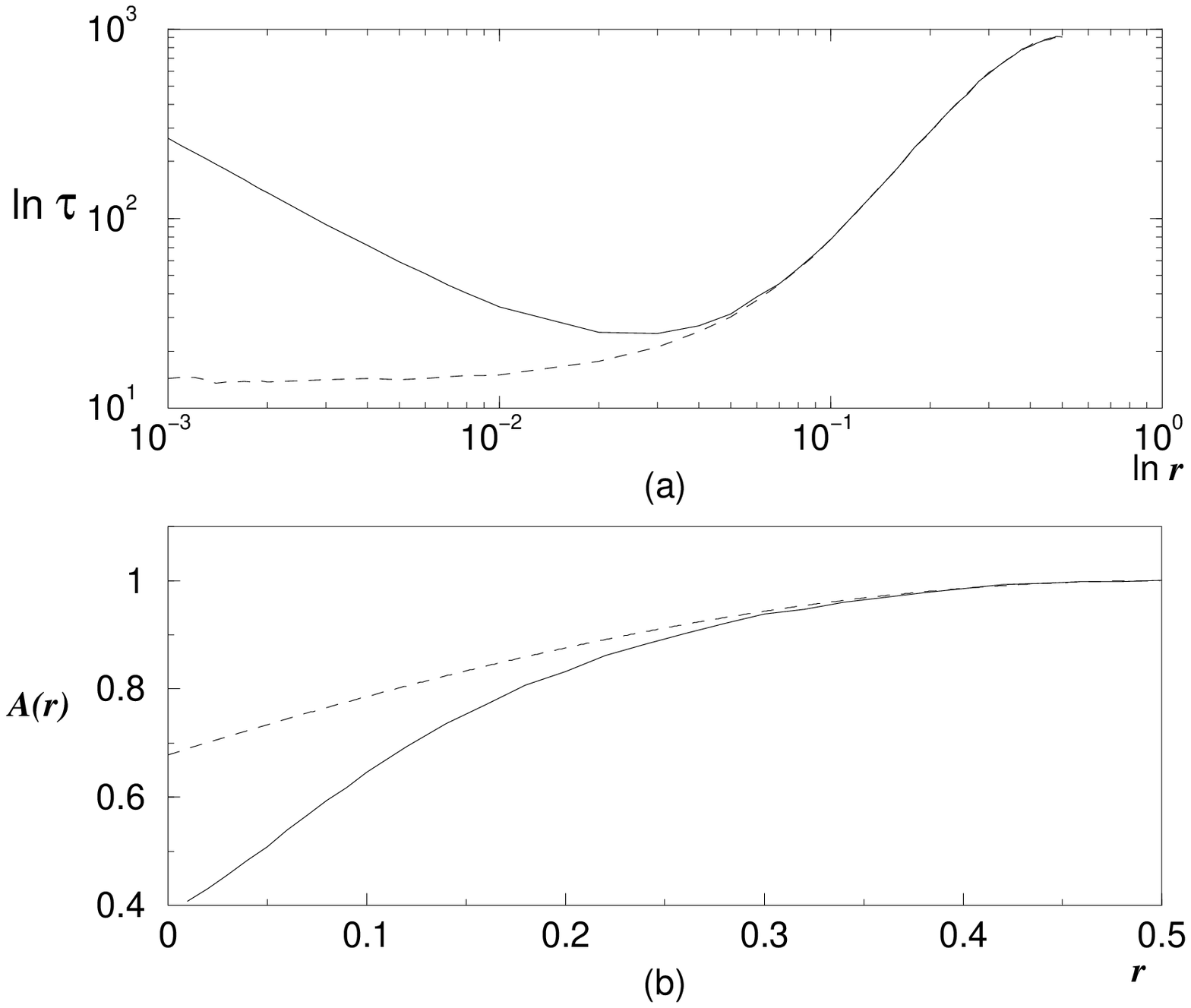,width=4in,clip=}
\centering
\caption{ \small  (a) Log-log plot of the crossing time $\tau$
as a function of the level of noise $r$. The solid curve is $\tau_d$,
the crossing time starting from different attractors, whereas the
dashed curve corresponds to $\tau_s$, the crossing time starting from
the same attractor. (b) The index $A(r)$ describing the fractal
dimension of the attractor plotted as a function of $r$. The solid
curve is the result of the numerical simulation, obtained as
$A(r)=2\ln\tau/\ln\Omega$, and the dashed line is the theoretical
prediction given in (\ref{A_r}). Note that these curves become
identical when $r\rightarrow0.5$. Both graphs (a) and (b), calculated
for a situation with $K=2$ and $N=20$, were taken from reference
\cite{Qu}.}
\label{fig:qu}
\end{figure}
%
%
%
%

The authors examine $K=4$ and large-$N$, picking the noise-levels $r=
0.15$ (low temperature phase), $r= 0.25$ (critical phase), and $r=0.35$
(high temperature phase). Without noise, the $N$-$K$ model would show
chaotic behavior, and very long cycles.  In the presence of noise the
average of $\ln \tau$, with $\tau$ being the convergence time,
diverges as $N$ goes to infinity.  The behavior of the convergence
time for two orbits, $\tau_2$, can be calculated in the annealed
sequential update case, and the results compared with simulations 
for both annealed and quenched system.  In both simulation and theory,
the three phases are characterized by having different forms of
divergence of $\langle\ln\tau_2\rangle$ with $N$ (see Fig.
\ref{Golinelli}).  At low temperatures the divergence is linear:
$\langle \ln\tau_2\rangle\sim N$. In the critical phase
$\langle\ln\tau_2\rangle$ diverges linearly with $\ln N$ as
$\langle\ln\tau_2\rangle=0.5 \ln N$.  The weakest divergence occurs in
the high temperature phase in which $\langle\ln\tau_2\rangle$ varies
as $\ln\ln N$.  These $N$-dependencies describe the variation of the
number of elements forming the barrier to having two configurations
merge into one another.  The noise causes these elements to be
identical and then the merge occurs. Thus, for example, in the low
temperature phase, we must bring to equality a finite fraction of all
the elements in the system in order to have the convergence.

In some loose sense, these numbers measure the size of the barriers
which hold together the attractors for this system.  But it is hard to
know what the attractors themselves might be.  Because $r$ is
relatively large, these presumed attractors are probably not the
cycles of the original system.  In fact, the behavior of $\tau_2$ is
much the same for the annealed system (which has no cycles) as for the
quenched system, which does have cycles at $r=0$.  Nobody has yet
investigated the limiting case as $r$ goes to zero.  It might be most
interesting to look at this limit, particularly in association with a
limit which keeps the system critical (say $K$ goes to 2).

Let us consider the case in which more than two trajectories are
analyzed. Start with $m$ trajectories and let $\tau_m$ measure the
first time when any two of these have converged.  If there are only
$M$ large or important basins, one might well expect
$\langle\ln\tau_m\rangle\ll\langle\ln\tau_M\rangle$ for $m>M$.
Instead one observes that
\[
\langle\ln\tau_m\rangle = \langle\ln\tau_2\rangle -
\ln \frac{m(m-1)}{2} ,
\]
in both theory (in the annealed approximation) and simulation.  This
form indicates an indefinitely large number of attractors, all with
basins of comparable size.

The crossing paper considers two different configurations $\Sigma_0^1$
and $\Sigma_0^{2}$, which can belong either to different attraction
basins or to the same basin of attraction, and then iterates forward,
noting all configurations $\{\Sigma_0^1$, $\Sigma_1^1,$
$\Sigma_2^1,\dots\}$ and $\{\Sigma_0^{2}$, $\Sigma_1^{2}$,
$\Sigma_2^{2},\dots\}$ they produce.  In each step of iteration, and
for each $i$-value, a choice is made between the two branches of
equation (\ref{m1cross}), and that choice is applied to both
configurations. This continues until the time $\tau$ in which one of
the two trajectories attains a configuration previously entered by the
other one (for example, if $\Sigma_\tau^{2}$ is equal to one of the
configurations $\{\Sigma_0^1$, $\Sigma_1^1,\dots$,
$\Sigma_\tau^1\}$). The calculation is terminated at this crossing
event. The measured quantity is the ``time'' $\tau$ needed to produce
the crossing.

In the absence of noise, if two initial configurations belong to
different basins of attraction, the time for the two subsequent
trajectories to cross is infinite.  In the presence of noise, there is
a chance for each trajectory to ``jump out'' of its basin of
attraction, exploring a bigger part of the state space.  The two
trajectories will have a number of opportunities equal to $\tau^2$, to
cross one another before $\tau$ steps have elapsed.  If the size of
the space being explored by the trajectories is $\Omega(r)$, then the
typical time for the crossing will be
\begin{equation}
\tau \approx \Omega(r)^{1/2} .
\label{crosstime}
\end{equation}
Miranda and Parga simulated the system and measured $\tau$ as a
function of $r$ and $N$ for the critical Kauffman net in which
$K=2$. Their result for large $N$ may be summarized as
\begin{equation}
\Omega(r) =\Omega^{A(r)} ,
\label{eq:Omega}
\end{equation}
where $\Omega=2^N$ is the volume of the state space in the system.

In the work by Miranda and Parga the ``fractal'' exponent $A(r)$ was
not estimated accurately.  This work was recently extended by X. Qu et
al., who consider larger values of noise and different connectivities
of the network (\cite{Qu}).  The authors analyze two cases to compute
the crossing time, when the two initial configurations belong to the
same basin of attraction, and when they belong to different basins. We
will denote these two crossing times by $\tau_s$ and $\tau_d$
respectively.  Fig. \ref{fig:qu}a shows the average crossing times
$\tau_d$ and $\tau_s$ as funcions of $r$, for a net with $N=20$ and
$K=2$. As can be seen, when $r$ is close to its maximum value $0.5$,
both times are practically the same. In fact, Qu et al. have shown
that for large values of $r$ both $\tau_s$ and $\tau_d$ behave as
\begin{equation}
\tau_{d,s}\approx\frac{\sqrt{\pi}}{2} 2^{N/2}
\left\{1 + \frac{(1-2r)^2}{2^K}\right\}^{-N/2}~.
\label{eq:tau_0.5}
\end{equation}
The above expression agrees with equations (\ref{crosstime}) and
(\ref{eq:Omega}) by identifying
\begin{equation}
A(r) = 1 - \frac{\ln\left[1+(1-2r)^2/2^K\right]}{\ln2}~.
\label{A_r}
\end{equation}

In contrast, when $r$ is close to $0$ the behavior of $\tau_d$ and
$\tau_s$ differ substantially. For $r\rightarrow0$ the divergence of
$\tau_d$ is given simply by
\begin{equation}
\tau_d\approx C_1/r + \tau_0
\label{eq:tau_0}
\end{equation}
where $C_1$ only depends on $K$ and $N$, and $\tau_0$ is the value of
$\tau_s$ at $r=0$.

The complete analytical expresion of $A(r)$, valid in the whole
interval $[0,1/2]$ is not known yet. Fig. \ref{fig:qu}b shows a plot
of $A(r)$ obtained by numerical simulations.  It is interesting to
note that the attractor has a fractal volume which depends upon $r$.
Once again, one is frustrated because one does not know what the
attractor might be.  It is once again probably not anything directly
related to a cycle, since starting points in the same cycle or in
different cycles both give the same $\tau$-values for the higher
values of $N$ and $r$.  Here too one might guess that studies with
smaller values of $r$ might shed light on the $N$-$K$ model
attractors.


\section{Applications}

\subsection{Genetic networks and cell differentiation}
\label{GenNet}

$N$-$K$ models have been widely used in the modeling of gene
regulation and control (\cite{96.6,97.6,93.1,00.4}). A very remarkable
characteristic of multicellular organisms is that all the different
cells of which they are made have the same genetic information.
What makes the difference between the different cell types are the
genes which are \emph{expressed} in every cell at every moment. In a
given cell type, some particular genes are turned off and others are
turned on. So, in a liver cell, only the ``liver genes'' are being
expressed while all the other genes are turned off, whereas in a
neuron the ``neuron genes'' are the only ones which are expressed.

\begin{figure}[h]
\centering
\psfig{file=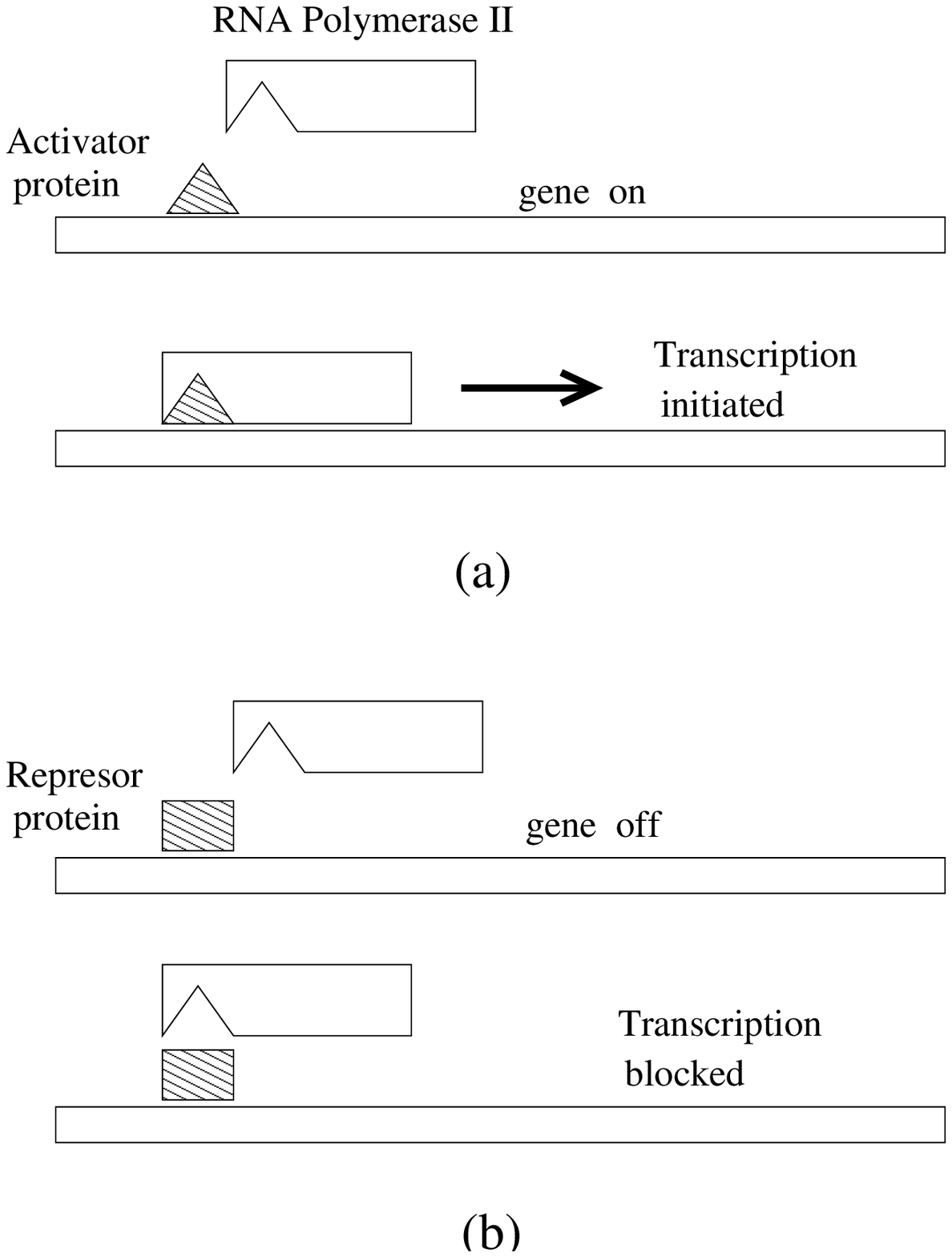,width=3in,clip=}
\caption[]{ \small Schematic representation of gene regulatory proteins.
  (a) An activator protein attaches to the gene at a specific binding
  site, activating the polymerase II which then can transcript the
  gene into a molecule of mRNA. (b) When a repressor protein is
  attached to the gene, the polymerase II is blocked up and therefore
  unable to transcript the information contained in the gene.}
\label{figa1}
\end{figure}

The physical and chemical mechanisms by which the cell determines
which genes are to be expressed and which are not are not yet fully
understood. There is evidence that gene regulation and control can
occur at every stage during the metabolic pathways leading up from the
genetic information contained in the DNA, to the translation of this
information into proteins. Nevertheless, most of the gene regulation
and control seems to occur at the level of transcription of the
genetic information.  At this level, one gene of DNA is transcribed
into a molecule of messenger ARN (mARN) only if the conditions for
this transcription are present. In the most simple model (applicable
to bacteria), for the transcription of one gene into a molecule of
mARN, it is necessary a protein, called \emph{activator}, which
attaches to the beginning of the gene indicating that this gene is
ready to be transcribed (see Fig. \ref{figa1}a). On the other hand,
there also exist \emph{repressor} proteins which, when attached to the
beginning of the gene, inhibit its transcription, turning the gene off
(see Fig. \ref{figa1}b).

In eucaryotic cells the situation is more complicated in that many
activator or repressor proteins might be needed to activate or to
inhibit the expression of a single gene. For example, it is known that
the human $\beta$-globine gene (expressed in red blood cells) is
regulated by more than 20 different proteins. Some of these proteins
may function as both repressors or activators, depending on how they
are assembled.

%
%
%
%
%
%
%
%

The activator or repressor proteins of a given gene are themselves
specified by some other genes, whose expression,  is in turn
controlled by other proteins codified by other genes and so on.  Genes
interact each other through the proteins they specify: the product
protein of one activated gene can influence the activation or
deactivation of other genes.  Similarly, the \emph{absence} of the
product protein of a deactivated gene can influence the
activation or deactivation of several other genes. In some particular
cases, it is known which gene controls which other one, but in most of
the cases the interactions between genes in a given cell type are
completely unknown. A picture emerges in which genes are mingled
together forming a network of interacting elements which are connected
in a very complicated way.

The $N$-$K$ model was first suggested by Kauffman as a way of modeling
the dynamics and evolution of this complicated network of interacting
genes (\cite{69.1}). Within this model, two genes are linked if the
product protein of one gene influences the expression of the other
one.  In real cells, genes are not randomly linked. Nevertheless, the
web of linkages and connectivities among the genes in a living
organism is so complicated (and mostly unknown), that to model these
genes as randomly linked seems to be a reasonable approximation. By
this means, the results coming out of the model are not restricted to
a particular set of linkages and connectivities, but reflect instead
the generic properties of this class of networks.

The state of the cell at every moment is determined then by the state
of its genes $\sigma_1,\dots,\sigma_N$. Before proceeding further, it
is important to recall the main assumptions that are usually made
to describe the dynamics of genetic networks by means of the $N$-$K$
model:

\begin{itemize}
\item Every gene $\sigma_i$ can be only in one of two states, turned
  on (1) or turned off (0).
\item Every gene is randomly linked with exactly $K$ other genes.
\item The evolution rule $f_i$ associated to gene $\sigma_i$ is a
  weighted function which acquires the value 1 with probability $p$
  ant the value 0 with probability $1-p$.
\item The updating of the network is synchronous.
\end{itemize}

As we have seen, under these assumptions the phase space breaks up
into different cycles or attractors whose properties in the frozen
phase and the chaotic phase are substantially different.  According to
Kauffman's interpretation, \emph{each attractor represents a cell type
or a cell fate}, whereas a single state of the system represents just
a temporal state which the cell is passing through.  From this point
of view, cell reproduction processes start with an initial
configuration of genes, which eventually evolves towards its
corresponding attractor; the attractor determines a particular cell
type or cell fate.

For this mechanism of cell differentiation to be meaningful, the
length of the attractors must not be too long, for otherwise the cell
would never reach its stable cycle. In the chaotic phase the length of
the cycles grows exponentially with the system's size ($l \sim
2^{\alpha N}$, where $\alpha$ is of order $1$). Therefore, the system
has to go through very many states before reaching a stable set of
configurations. In addition, systems in the chaotic phase are very
sensitive to perturbations (mutations), partially because the number
of relevant elements is comparable to the size of the system.  The
above prompts the thought that genetic networks of living organisms
\emph{are not} in the chaotic phase.

On the other hand, in the frozen phase the cycles
are much shorter than in the chaotic phase: an initial
set of genes swiftly reaches its stable configurations.  The
fraction of relevant elements in the frozen phase is close to zero.
As a consequence, the system is extremely resistant to point mutations
\footnote{A point mutation is a change in the value of one gene.} or to
damage in one or more of the evolution rules.  But in order to evolve,
genetic networks of living organisms should allow some degree of
sensitivity to mutations, which rules out the frozen phase as a
physical state which living organisms could be in.

Kauffman suggested that gene networks of living organisms operate
\emph{at the edge of chaos} (\cite{93.1}),
meaning that the parameters have been adjusted through evolution so
that these networks are at or near the critical phase. In the critical
phase, both the number of different attractors as well as their
lengths are proportional to a power of $N$, so the cell can reach very
quickly its stable configurations. Also, the fraction of relevant
nodes in critical networks, even though is small, is not zero, which
means that these kinds of networks present some degree of sensitivity
to changes in the initial conditions. In other words, critical
networks exhibit \emph{homeostatic stability}, a term which we will
come to in the next section.

A very remarkable observation supporting the idea of life at the edge
of chaos consists in the fact that the number of different cell types
in an organism is roughly proportional to the square root of its DNA
content.  Furthermore, the mitotic cycle period, which can be
considered as a measure of the time required for a cell to reproduce,
seems also to be proportional to the square root of the cell's DNA
content (\cite{93.1}).  Thus, random networks in the critical phase
seem to satisfy the requirements of order, evolvability and stability
found in living organisms.

Even though this idea is very attractive, there are some problems yet
to be solved.  As we have mentioned, for unbiased evolution rules
$f_i$, the critical phase is characterized by the low connectivity
$K=2$. This implies that genetic networks of real organisms are
restricted to have very low connectivities in order to be at ``the
edge of chaos''. As soon as the connectivity grows, the system becomes
more and more chaotic. But it is well known that the connectivity in
real genetic networks is rather high. For example, the expression of
the \emph{even-skipped} gene in \emph{Drosophila} is controlled by
more than 20 regulatory proteins, also the Human $\beta$-globine gene
we have referred to before (\cite{94.4}). In eucaryotes it is common
to find that one single gene is regulated by a bunch of proteins
acting in association.  On the other hand, sometimes when a single
signaling receptor protein is activated, it can influence directly the
activation of a very large array of genes. Let us consider for
instance the activation of the platelet-derived growth factor $\beta$
receptor (PDGFR$\beta$), which induces the expression of over 60 genes
(\cite{99.17}).  These examples, among many others, suggest that the
connectivity in real genetic networks is not low, but on the contrary,
it is very high.  Nonetheless, cells do not seem to operate in the
chaotic phase.

There are two ways to increase the connectivity in the $N$-$K$ model
without going out of the critical or ordered phase: (a) by the use of
weighted evolution functions $f_i$, or (b) by the use of canalizing
functions.  When evolution functions $f_i$ are weighted with the
probability parameter $p$, the critical line is given by equation
(\ref{K_c}).  Fig. \ref{critical_line} shows the graph of $K_c$ as a
function of $p$. Even if we suppose that genetic networks can be
either in the frozen phase or in the critical phase, they would be
restricted to remain within the shaded area of the figure.

On the other hand, it is known that the fraction of canalizing
functions becomes very small as $K$ increases (\cite{84.1}).
An upper bound for this fraction is
\begin{equation}
\frac{4K}{2^{2^{K-1}}} ,
\label{canalizing}
\end{equation}
which tends to zero as $K\rightarrow \infty$. Consequently, canalizing
functions are extremely rare when $K$ is large\footnote{Even for $K$
as small as $K=10$, equation (\ref{canalizing}) gives a fraction of
canalizing functions of the order $10^{-153}$.}. If the apparent order
seen in living cells relies on either weighted or canalizing
functions, it has still to be solved what kind of mechanisms drove,
through evolution, the genetic networks towards the generalized use of
such type of functions.

Another problem lies in the assumption that genes can be in only two
states. It is true that a given gene is expressed or is not.  But the
product protein of a gene can participate in a variety of metabolic
functions, producing even opposite effects depending on the physical
and chemical context in which it acts. Such behavior can be modeled by
assuming that every gene can acquire more than two states.  But in
such a case the connectivity of the network must be even smaller to
keep the system within the ordered phase. For if we assume that every
gene can be, on average, in one of $m$ possible states, and if every
one of these states is activated with the same probability, the
critical connectivity $K_c$ is then given by
\begin{equation}
K_c = \frac{m}{m-1} .
\end{equation}
The critical connectivity decreases monotonically when $m > 2$,
approaching $1$ as $m \rightarrow \infty$.  The moral is that for this
kind of multi-state networks to be in the ordered phase, the
connectivity has to be very small, contrary to what is observed in
real genetic networks.

Partially to overcome these difficulties, \cite{97.6} have proposed a
model which is slightly different from the original $N$-$K$ model.  In
their model, the dynamics of the system is governed by the equation
\begin{equation}
\sigma_i(t+1)=\mbox{Sign}\left\{J_{ii}\sigma_i(t)+\sum_{l=1}^{K-1}
J_{ij_l(i)}\sigma_{j_l(i)}(t)\right\} ,
\end{equation}

%
%
%
%
\begin{figure}[]
\centering
\psfig{file=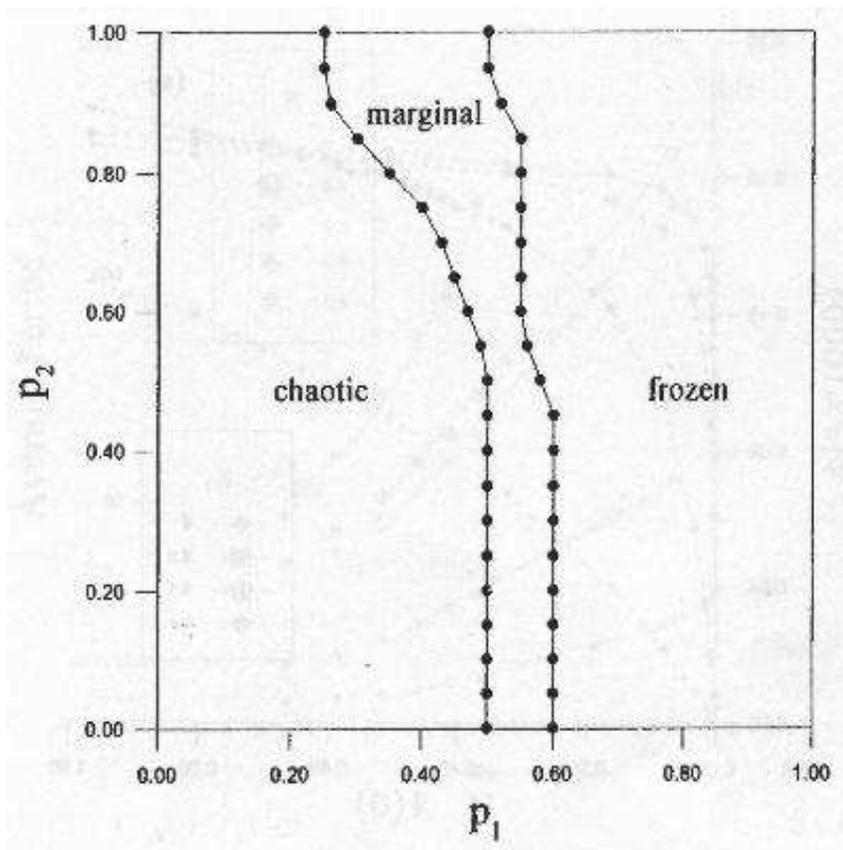,width=\hsize,clip=}
\caption[]{ \small Phase space for the $N$-$K$ model with the dynamics
  given by equation (\ref{nkdynamics}). As can be seen, under this
  dynamics the system exhibits three different ``phases'': frozen,
  marginal and chaotic (taken from  \cite{97.6}).}
\label{figa3}
\end{figure}
%
%
%
%

\noindent where $J_{ij_l(i)}$ is the coupling constant representing the
regulatory action of the $j_l(i)$ input of gene $i$, $(l = 1,
2,\dots,K-1)$.  There is also an autogenic regulation, given by
$J_{ii}$. The set of coupling constants $J_{ij}$ represents the very
complicated set of biochemical interactions between genes. Since these
interactions are mostly unknown, the authors assign the coupling
constants in a random way, according to the following criteria:

\begin{itemize}
\item The product protein of a given gene can activate, inhibit, or
  not affect at all the transcription of another gene.
\item Interactions between genes are not necessarily symmetrical, namely,
  $J_{ij}\neq J_{ji}$ in general.
\item Autogenic or self-regulation is allowed.
\end{itemize}

The coupling constants are then assigned according to the probability
function given by:
\begin{equation}
P(J_{ij})=\frac{1-p_1}{2}\left[\delta(J_{ij}-1)+\delta(J_{ij}-1)\right]
+ p_1\delta(J_{ij}) .
\label{nkdynamics}
\end{equation}
In this way, the couplings $J_{ij}$ can be activating ($+1$) or
inhibitory ($-1$), each with probability $(1-p_1)/2$, or neutral (0)
with probability $p_1$. Furthermore, the linkages for every gene are
chosen either at random among the whole set of genes, with probability
$p_2$, or only among the nearest neighbors, with probability
$1-p_2$. By varying the parameter $p_2$ one can go from lattices where
local interactions are the most important, to random nets where all
ranges of interaction are present.

By analyzing the Hamming distance between two initial configurations,
the authors show that there are three phases in the dynamics of the
system: a frozen phase, a marginal phase and a chaotic phase, as shown
in Fig. \ref{figa3}.  The frozen and chaotic phases are as in the
traditional $N$-$K$ model. The marginal phase is characterized by the
fact that all the attractors are stable since a small change in the
initial configuration neither vanishes nor grows. Also, in this phase
both the number of different attractors and their lengths grow as a
power of $N$.  The above results were obtained for networks with
connectivity $K=9$ and $N=625$. It is probable that the marginal phase
is only a finite-size effect, vanishing for an infinite system. More
work has to be done to explore the whole space of parameters of this
model, but these preliminary results show that it is possible to
obtain ordered and marginal behaviors even in the case of high
connectivities by changing the type of interaction between the genes.

\subsection{Evolution}  

In a traditional framework of the theory of evolution
(\cite{69.3,89.8,86.5}), changes in the phenotype of organisms are
accumulated gradually, yielding a gradual increase in complexity of
form and function.  Simple structures slowly assemble together to form
more complex structures, which in turn assemble to build up even more
elaborate systems, and so on.  At every stage in the formation of a
complex system (organism) out of simpler elements, many sub-systems
are created, which represent temporal stable states along the way in
the construction of the whole system. As T. D. Zawidzki has pointed
out in \cite{98.10}, from this point of view the problem of evolution
of complex systems translates into a search problem. For the
construction of a complex system, evolution searches in the space of
the possible configurations leading to the whole system. If no stable
intermediate configurations were found in this search, the
construction of the whole system would be extremely improbable, since
the number of possible configurations in which the parts can be
arranged increases exponentially with the number of different parts.
Evolution does not search in one step for the ``final'' configuration
of a complex system. Instead, the search is carried out gradually,
finding first intermediate stable configurations of sub-systems that
are then assembled, giving rise to the whole complex system. Every
sub-system solves a particular problem (or set of problems),
facilitating the construction and functioning of the whole organism.
Furthermore, once an evolutionary problem has been solved by primitive
organisms, more complex species which evolve from them still continue
solving this problem in the same way.  New species are faced with new
problems, but still preserving the old solutions to the old problems
(the genetic code, for example, was ``invented'' only once).
According to this thesis, genomes of complex organisms are made up of
functional modules of genes, each module encoding the solution of a
given biological problem encountered by the species at some point
through evolution.

The study of Boolean networks has suggested new mechanisms for
evolutionary processes.  Kauffman has stressed that many evolutionary
changes involve reorganizing of the same genetic material rather than
making it more complex (\cite{93.1,95.4}).  In Kauffman's approach,
genes are organized in complex genetic networks provided with some
given dynamics. The ``search'' of evolution consists in searching for
the most stable organization of genes, which in terms of the $N$-$K$
model means stable cycles. From this point of view, there is not an
increasing complexification of the evolving system, but a inherent
complex organizational dynamics which settles down in a finite number
of stable attractors. The ``role'' of evolution then is to look for
the more stable attractors which the system can fall into.

A Boolean network made up of $N$ genes has $2^N$ states, but the
system organizes itself into a much smaller number of attractors.
Depending on the parameter values being used ($K$ and $p$), these
attractors are stable or they are not. Stability is defined according
to the response of the network to perturbations, which can be of three
different kinds:

\begin{itemize}
\item changes in the states of a few genes by flipping the value of
  some randomly chosen ones;
\item permanent changes in the linkages of some genes;
\item permanent changes in the values of the evolution functions $f_i$
  associated with some genes.
\end{itemize}

As we have already mentioned, only networks in the critical phase have
the stability required to constitute evolvable systems, in that these
networks are able to recover to most of the mutations described above.
In the chaotic phase, the attractors are extremely unstable since any
kind of perturbation would shift the system to another attractor. In
the frozen phase, even though the majority of the genes are
motionless, small changes in the linkages or in the evolution
functions would make the system jump to a very different attractor
if these damages are carried out on the relevant elements of the net.
But networks operating in the critical phase show a very high
homeostatic stability, which means that after some perturbation, such
networks are very likely to fall again in the same attractor.
Consequently, in this approach, evolution is also interpreted as a
sort of search for gene networks possessing stable dynamics and not
merely as a searching for stable sub-systems out of which more
complicated systems can be built up. The fact that more complex living
organisms have bigger amounts of genetic material, together with the
fact that real genetic networks actually exhibit high homeostatic
stability, suggest that evolutionary processes consist of both kinds
of ``searching,'' hierarchical-modular complexification and dynamical
stability.

So far we have considered processes occurring at the level of genomes,
but evolution also acts at the level of populations, making the
organisms of a given species become better adapted to their
environment. Living beings are subjected to all kinds of external
fluctuations, and the survival of the species depends on the
capability of its members to recover from those random
perturbations. Evolutionary processes produce organisms with a high
degree of homeostatic stability and of adaptability to the
environment, even in the presence of external fluctuations. When
thinking of evolutionary processes, one usually supposes that living
organisms tend to exclude noise since stability is more conductive to
functioning than chaos, and also that evolutionary processes are able
to recognize and favor such stability.  Nevertheless, Michael D. Stern
has pointed out that none of these assumptions has been rigorously
proven yet (\cite{99.9}). Furthermore, he has shown that, under
certain circumstances, noise not only is not excluded from an
adaptative system, but it is \emph{required} for the adaptation of the
system to the environment; without noise, such adaptation would not be
possible.

In this work Stern considers a population of $M$ organisms, each of
which is a $N$-$K$ model composed of $N=100$ elements. Each organism
is in the critical phase ($K=2$ and $p=0.5$) and therefore, according
to Kauffman, they are in the state of highest homeostatic stability.
The linkages among the elements and the evolution rules $f_i$ are
assigned in the usual way, but additionally, a noisy signal $\eta(t)$
is applied to $m$ randomly chosen elements of every organism. The
noise is introduced through only one of the two inputs of each one of
the $m$ elements.  Suppose for example that $\sigma_{i_1}$ and
$\sigma_{i_2}$ are the two inputs of $\sigma_i$. With no noise, the
value of $\sigma_i$ at time $t+1$ would be given, as usual, by
\[
\sigma_i(t+1)=f_i\left(\sigma_{i_1}(t),\sigma_{i_2}(t)\right) .
\]
But if $\sigma_i$ is one of the $m$ elements to which the noisy signal
$\eta(t)$ is being applied, then the value of $\sigma_i$ at time $t+1$
is given now by
\[
\sigma_i(t+1)=f_i\left(\sigma_{i_1}(t),\eta(t)\right) .
\]
Note that the noisy signal $\eta(t)$ is the same for the $m$ elements
of every one of the $M$ organisms. This takes into account the fact
that in a real population, every organism in the population is
subjected to the same (noisy) environment.

The evolution of the population is now determined by an external
criteria which has to be fulfilled. This is the \emph{phenotype} of
the population on which natural selection will be acting.  In Stern's
work, the phenotype to be selected is an integer time series, obtained
by counting the number of positive states ($+1$) occurring in a given
subset of elements (output elements) in each organism. This time
series is to fit a predefined time function $F(t)$ (the target
function), and in each generation the organisms better adapted to
$F(t)$ are selected.

Selection of the organisms is made as follows. At the beginning of
each generation, every organism is replicated $R$ times allowing some
mutations which consist mainly of randomly moving one of the input
connections of a randomly chosen element, and randomly changing the
Boolean function of a randomly chosen element. So, at the beginning of
every generation the population actually consists of $R\times M$
elements. The system is then evolved during 100 time steps, after
which the $M$ elements which best fit the external criteria $F(t)$ are
selected, starting another generation.

%
%
%
%
\begin{figure}[]
\centering
\psfig{file=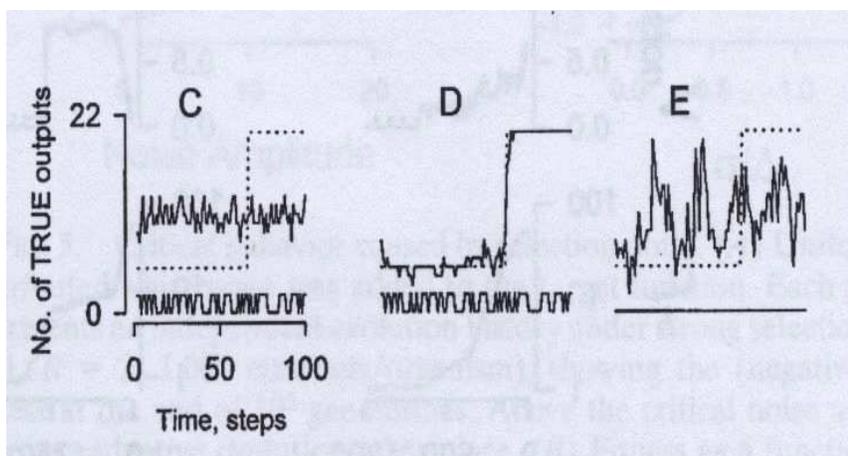,width=\hsize,clip=}
\caption[]{ \small (C) Random output of the starting network (solid),
  the target function (dotted), and the fixed binary noise added to
  the system (lower) in Stern's model.  (D) The evolved network
  generates a good approximation of the target function after 2500
  generations. (E) The same network as in D but operating in the
  absence of noise fails completely in fitting the target function,
  indicating that the evolved network has been ``imprinted'' by the
  arbitrary noise sequence present during its evolution and requires
  it to function (taken from \cite{99.9}).}
\label{figa4}
\end{figure}
%
%
%
%

The results of the simulation are shown in Fig. \ref{figa4}, in which
three output series are compared with the target function $F(t)$
(dotted line). Fig. \ref{figa4}C shows the output signal of the
initial generation, which has not yet passed through any selection
process.  In the presence of noise $\eta(t)$, after 2500 generations
the target function is very well approximated by the output signal of
the evolving network, as shown if Fig. \ref{figa4}D. However, in the
absence of noise the same network fails completely to approximate the
target function (Fig. \ref{figa4}E). This surprising result implies
that noisy perturbations may be essential for the adaptation of
organisms to their environments. Stern has called this phenomenon
\emph{noise imprinting} in evolution, and claims that it ``may be a
prototype of a general form of symmetry breaking that leads to the
evolution of structures that are suboptimal and unnecessarily
dependent on fortuitous features of the environment''.

\subsection{Social Systems}

It is always difficult to extrapolate the models and techniques of
physics to the realm of social sciences.  To begin with, it is not yet
clear how to define a social system in a way that is suitable for the
kind of analysis that physicists are used to. The complexity we see in
human societies is of a quite different nature than the one present in
physical systems, so one must be extremely cautious when talking about
``complexity'' and ``chaos'' in social sciences. Nonetheless, some
parallels can be established between the complex behaviors of Boolean
networks and human organizations.

Boolean networks were first introduced in social sciences by Schelling
in 1971 (\cite{71.1}).  Following
\cite{99.14}, the main aspect in a social system
is that \emph{social actors act according to certain social rules; the
intended or unintended consequences of their actions generate other
actions by the same or other actors and so forth}.
The result of such
interactions is reflected in the dynamics of the system.

A key difficulty in the modeling of social networks is determining the
quantities that convey enough information to describe a given
system, and which have some degree of predictability.  One must also
identify the key parameters governing the behavior.  Two kind of
parameters can be distinguished: those which determine the topology of
the system, namely, how individuals interact, and the ones which
determine the social rules, i.e., the consequences of the
interactions. To illustrate these two kind of parameters, we can focus
our attention on the connectivity $K$ of Boolean networks and the
weight parameter $p$ associated to the evolution functions $f_i$.

In ancient societies, characterized by very rigid dictatorial or
monarchic regimes, the connectivity among people was very small since
the majority of individuals were restricted to interact only with the
few people in their communities.  Mobility of individuals among
different social classes was also minimal. Poor people were condemned
to remain poor while rich people usually remained rich; social
mobility, if any, was allow only from upper to lower classes. The
course of the whole society depended largely on very few people (the
King and his clique). In contrast, modern democratic societies are
characterized by much bigger connectivities since individuals have a
broader spectrum of interactions. In addition, social mobility is
greater in democratic societies since, at least in principle, anybody
can become president or actor or any other thing.

A binary network can be used for the modeling of a society of
interacting individuals, in which poverty is represented by 1 while
richness is represented by 0. Of course, in real societies the
connectivity among individuals varies from one individual to another,
and so does the range of influence of their decisions; but it is clear
that even if we assign the same connectivity to each member of the
society, the out-degree will not be the same for all
individuals\footnote{The out-degree of node $\sigma_i$ is the number
of different nodes which are affected by $\sigma_i$, and can run from
0 up to $N$.}, and therefore the range of influence that each member
of the society has on the other members varies among them.

With this setup, a rigid feudal society would be characterized by low
values of the parameter $K$ and by evolution functions $f_i$ whose
weight parameter $p$ is very low, reflecting the fact that the
majority of the individuals are poor no matter what they do or whom
they interact with. This kind of society presents a very simple
dynamics in a frozen regime; no changes are expected nor unpredictable
behavior. The dynamics is governed by a very reduced fraction of
relevant elements, while the rest of the individuals remain in a
frozen state with nearly zero mobility.
 
On the other hand, a perfect democratic society would be rather
characterized by high connectivities and weighted functions $f_i$
whose parameter $p$ is close to $0.5$. In this situations, we have
seen that the system exhibits chaotic behavior, with an apparently
random dynamics. The society would never reach a stable state or a
stable attractor, for in this regime the length of the cycles is
extremely large.

If these results reflect some of the fundamental aspects behind the
dynamics of social organizations, the conclusion would be that the
complexity that we see in modern democratic societies is inherent to
the democratic principles (parameter values) on which these societies
have been constructed. Apparently, democracy and complexity are tied
together, and it is a political decision whether or not it is
worthwhile to sacrifice democracy in the interest of
predictability. We should stress, though, that Boolean networks are
far from accurate representations of a human society, so we should
consider the previous results only as indicative trends and not as the
matter of the facts.

\subsection{Neural Networks}

The first neural network model was introduced nearly 60 years ago by
McCulloch and Pitts in an attempt to understand some of the cognitive
processes of the brain (\cite{43.1}). The topic of neural networks has
since grown enormously, and presently it covers a great variety of
fields, ranging from neurophysiology to computational algorithms for
pattern recognition and non-parametric optimization processes
(\cite{94.5}). Nonetheless, it is somehow accepted that the two major
applications of neural networks consist in the understanding of real
neural systems (the brain), and in the development of machine-learning
algorithms (\cite{90.5}). In this section, we will address the subject
from a point of view based on statistical physics.

As in the $N$-$K$ model, a typical neural network consists of a set of
$N$ nodes, each of which receives inputs from $K$ other nodes.  The
difference from the $N$-$K$ model is that in a neural network the
dynamics are given by\footnote{There are other choices for the
dynamics, such as
\[
\sigma_i(t+1)=\tanh\left(\sum_{j=1}^Kw_{ij}\sigma_{i_j}(t) +
h\right) ,
\]
which produces an output between $-1$ and $+1$, or
\[
\sigma_i(t+1)=\left(1+\exp\left\{\sum_{j=1}^Kw_{ij}\sigma_{i_j}(t) +
h\right\}\right)^{-1} ,
\]
with an output between 0 and 1. Possible choices abound, depending on
the particular task for which the network has been designed. We will
focus on the particular class of neural networks obeying the dynamics
given by equation
\ref{neural_network}.}
\begin{equation}
\sigma_i(t+1)=\mbox{Sign}\left(\sum_{j=1}^Kw_{ij}\sigma_{i_j}(t) + h\right) , 
\label{neural_network}
\end{equation}
where the \emph{synapse weights} $\{w_{ij}\}$ and the
\emph{activation threshold} $h$ are random variables that can adapt
(in learning processes) or stay fixed (in recall or association
processes). Note that with the above definitions, the nodes have the
values $\sigma_i=\pm 1$.

%
%
%
\begin{figure}[]
\centering
\psfig{file=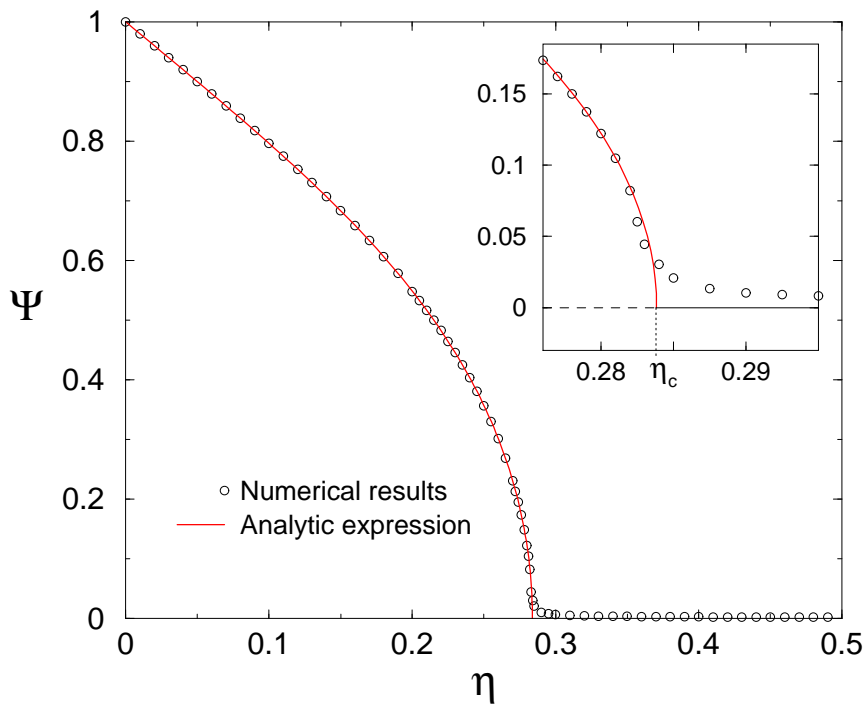,width=\hsize,clip=}
\caption[]{ \small Bifurcation diagram of the order parameter
$\Psi$ as a function of the noise intensity $\eta$, for the neural
network model in (\ref{CM_model}). In the case shown, the network was
composed of $N=10^5$ elements with connectivity $K=11$, and the
synapse weights $w_{ij}$ were uniformly distributed between 0 and
1. The inset shows the finite-size effects in the numerical results
near the critical noise $\eta_c\simeq 0.2838$. (Taken from
\cite{02.1}.)}
\label{figCMmodel}
\end{figure}
%
%
%

K. E. K\"urten has considered the case in which the synapse weights
are independent random variables distributed according to a symmetric
probability function $\mathrm{P}(w)$. By analyzing the Hamming
distance, he has shown that this kind of neural network has behavior
similar to that of the $N$-$K$ model, in that the three phases
(frozen, critical and chaotic) are also present (\cite{88.11,88.13}),
depending on the connectivity $K$ and the dilution of the network
(dilution is a measure of the amount of nodes for which the synapse
weight is 0). The main result is that for low connectivity and
dilution, the neural network and the $N$-$K$ model exhibit exactly the
same dynamics in the Hamming distance.

Another important application of neural networks consists in
investigating the tolerance to the influence of external noise in the
dynamical organization of the network. \cite{02.1} have analyzed a
neural network whose dynamics is given by
\begin{equation}
\sigma_i(t+1) = \left\{
\begin{array}{lcc}
\mbox{Sign}\left(\sum_{j=1}^Kw_{ij}\sigma_{i_j}(t)\right) &
\mbox{with probability} & 1-\eta , \\
& & \\
-\mbox{Sign}\left(\sum_{j=1}^Kw_{ij}\sigma_{i_j}(t)\right) &
\mbox{with probability} & \eta .
\end{array}
\right.
\label{CM_model}
\end{equation}
The dynamics can thus be changed from purely deterministic to purely
random by varying the noise intensity $\eta$ between 0 and 1/2. In the
case $\eta=0$ the network will typically converge to an ordered state
in which all the $\sigma_i$ acquire the same value. For other values
of $\eta$, the instantaneous amount of order in the networks is
determined by the parameter
\begin{equation}
s(t)=\lim_{N\rightarrow\infty}\frac{1}{N}\sum_{i=1}^N\sigma_i(t) ,
\end{equation}
so that $|s(t)|\approx 1$ for an ``ordered'' system in which most
elements take the same value, while $|s(t)|\approx 0$ for a
``disordered'' system in which the elements randomly take values $+1$
or $-1$. The order parameter of the network is defined then as the
time-independent quantity
\begin{equation}
\Psi=\lim_{T\rightarrow\infty}\frac{1}{T}\int_{0}^T|s(t)|dt .
\label{order_parameter}
\end{equation}
Huepe and Aldana have shown that, under very general conditions, the
neural network model given in (\ref{CM_model}) undergoes a dynamical
second order phase transition, as it is illustrated in
Fig. \ref{figCMmodel}.  More specifically, if the probability density
$\mathrm{P}(w)$ of the synapse weights is a non-symmetric, but
otherwise arbitrary function, and if the linkages of the network are
chosen randomly, then there exists a critical value $\eta_c$ of the
noise in the vicinity of which the order parameter behaves as
\begin{equation}
\Psi =
\left\{
\begin{array}{lcc}
C\left(\eta_c-\eta\right)^{1/2} & \mbox{for} & \eta<\eta_c ,\\
 & & \\
0 & \mbox{for} & \eta > \eta_c ,
\end{array}
\right. 
\end{equation}
where $C$ and $\eta_c$ are constants whose values depend on the
connectivity $K$ and on the particular form of $\mathrm{P}(w)$.  It is
worth mentioning that the dynamics of the network are exactly the same
if both the linkages and the synapse weights $w_{ij}$ are
time-independent, or if they are re-assigned at every time step.  As
far as we know, this is the first network model in which it has been
shown that the quenched and annealed approaches coincide exactly.

Other models of particular interest are those in which the synapse
weights $w_{ij}$ are in turn given by ``sub-layers'' of connections,
such as in the Little-Hopfield model (\cite{82.1,74.2}),
\[
w_{ij}= C_{ij}\sum_{\mu=1}^p\xi_i^\mu\xi_j^\mu ,
\]
where $\xi^\mu_i=\pm 1$ is the value of site $i$ in the sub-layer
$\mu$, and the coefficients $C_{ij}$ are independent random variables
distributed according to some given probability distribution
$\mathrm{P}(C)$.  These networks are important because by suitable
changing the coefficients $C_{ij}$, the network can be either
synchronized or ``taught'' some particular task (\cite{99.18,96.7,87.4}).

\section*{Acknowledgements}
We would like to thank Xiaohui Qu for her valuable help in the
elaboration of this article. This work was partially supported by the
MRSEC Program of the NSF under Award Number 9808595, and by the
NSF-DMR 0094569. We also thank the Santa Fe Institute of Complex
Systems for partial support through the David and Lucile Packard
Foundation Program in the Study of Robustness. M. Aldana also
acknowledges CONACyT-M\'exico for a postdoctoral grant.

\bibliographystyle{marsden}
\bibliography{bibliography}

\end{document}